\documentstyle[12pt,graphicx,amsmath,amssymb]{article}
\newdimen\singlebaseskip		
\newdimen\doublebaseskip		
\font\eightrm=cmr8			

\font\bbbrm=cmbx10 scaled\magstep1
\font\bbrm=cmbx10

\font\tenbmit=cmmib10		
\font\sevenbmit=cmmib10 at 7pt	
\font\fivebmit=cmmib10 at 5pt	
\mathchardef\BOLDalpha="710B 	
 \mathchardef\BOLDbeta="710C 	
\mathchardef\BOLDgamma="710D 	
  \mathchardef\BOLDrho="711A 	
\def\={\overline}		
\def\cms{\ifmmode {\rm\,cm\,\,s^{-1}} \else {$\cms$} \fi} 
\def\cmss{\ifmmode {\rm\,cm\,\,s^{-2}} \else {$\cmss$} \fi} 
\def\deg{\ifmmode {^\circ} \else {${}^\circ$} \fi}	
\def\etal{{\it et al. \/}}	
\def\gcms{\ifmmode {\rm\,g\,\,cm^{-2}} \else {$\gcms$} \fi} 
\def\gcmc{\ifmmode {\rm\,g\,\,cm^{-3}} \else {$\gcmc$} \fi} 
\def\kms{\ifmmode {\rm\,km\,\,s^{-1}} \else {$\kms$} \fi} 
		% ...
\def\lsun{\ifmmode {\rm\,L_\odot} \else {$\lsun$} \fi}	
\def\msun{\ifmmode {\rm\,M_\odot} \else {$\msun$} \fi}	
\def\rsun{\ifmmode {\rm\,R_\odot} \else {$\rsun$} \fi}	
\def\s{\ifmmode \widetilde \else \~\fi} 
\def\spose#1{\hbox to 0pt{#1\hss}} %
\def\Dt{\spose{\raise 1.5ex\hbox{\hskip3pt$\mathchar"201$}}}    
\def\dt{\spose{\raise 1.0ex\hbox{\hskip2pt$\mathchar"201$}}}    
\def\gta{\mathrel{\spose{\lower 3pt\hbox{$\mathchar"218$}}
     \raise 2.0pt\hbox{$\mathchar"13E$}}}
\def\lta{\mathrel{\spose{\lower 3pt\hbox{$\mathchar"218$}}
     \raise 2.0pt\hbox{$\mathchar"13C$}}}

\newcount\notenumber\notenumber=1
\def\foot#1{\raise3pt\hbox{\eightrm \the\notenumber}
     \hfil\par\vskip3pt\hrule\vskip6pt
     \noindent\raise3pt\hbox{\eightrm \the\notenumber}
     #1\par\vskip6pt\hrule\vskip3pt\noindent\global\advance\notenumber by 1}
\def\note#1{\footnote{$^{\the\notenumber}$}{#1}\global\advance\notenumber by 1}
\def\alph#1{\ifcase#1\or a\or b\or c\or d\or e\or f\or g\or h\or i\or j\or
	k\or l\or m\or n\or o\or p\or q\or s\or t\or u\or v\or w\or x\or
	y\or z\else #1\fi}
\def\Alph#1{\ifcase#1\or A\or B\or C\or D\or E\or F\or G\or H\or I\or J\or
	K\or L\or M\or N\or O\or P\or Q\or S\or T\or U\or V\or W\or X\or
	Y\or Z\else #1\fi}

\def\Roman#1{\expandafter\uppercase\expandafter{\romannumeral #1}}
\newcount\sectno    \sectno=0
\newcount\subno     \subno=0
\newcount\subsubno  \subsubno=0
\newcount\eqnmbrsec \eqnmbrsec=0
\newcount\eqnmbr    \eqnmbr=0
\def\sectionbeg#1{\penalty-200\bigskip\par 
	\advance\sectno by 1
	\subno=0\subsubno=0\eqnmbrsec=0
	\noindent{\bbbrm \Roman\sectno. #1}
	\nobreak\smallskip\par}

\def\subsectbeg#1{\penalty-200\medskip\par
	\advance\subno by 1\subsubno=0
	\noindent\hskip 0.25truein {\bbrm \Alph\subno. #1}
	\nobreak\smallskip\par}
\def\subsubbeg#1{\penalty-200\medskip\par
	\advance\subsubno by 1
	\noindent\hskip 0.50truein {\it \number\subsubno. #1}
	\nobreak\smallskip\par}

\def\eqnam#1{\xdef#1{\the\eqnmbr}}		  
\def\eqnamA#1{\xdef#1{{\it A}\the\eqnmbr}}		  
\def\eqnumrun{\global\advance\eqnmbr by 1 \the\eqnmbr}
\def\eqnumsec{\global\advance\eqnmbrsec by 1 \the\sectno.\the\eqnmbrsec}

\begin{document}
\begin{titlepage}
      \vspace{2\baselineskip}
      \vspace{2\baselineskip}
      \vspace{2\baselineskip}
\begin{center}
\begin{large}
      {\bf Deciphering the Origin of the Regular Satellites of Gaseous Giants -- Iapetus: the Rosetta Ice-Moon}\\
      \vspace{2\baselineskip}
Ignacio Mosqueira and Paul R. Estrada \\
Carl Sagan Center, SETI Institute\\
and\\
Sebastien Charnoz\\
Universit\'{e} Paris Diderot/CEA/CNRS\\
      August 2009\\
\end{large}
\end{center}
\vspace{1\baselineskip}
\begin{center}
{\bf Abstract}
\end{center}

Ever since their discovery the regular satellites of Jupiter
and Saturn have held out the promise of providing
an independent set of observations with which 
to test theories of planet formation. Yet elucidating 
their origin has proven elusive. Here we show that 
Iapetus can serve to discriminate between satellite
formation models. Its accretion history can be understood
in terms of a two-component gaseous subnebula,
with a relatively dense inner region, and an extended tail
out to the location of the irregular satellites, as in the 
SEMM model of Mosqueira and Estrada (2003a,b). 
Following giant planet formation, planetesimals in the 
feeding zone of Jupiter and Saturn become dynamically 
excited, and undergo a collisional cascade. 
Ablation and capture of planetesimal fragments
crossing the gaseous circumplanetary disks delivers 
enough collisional rubble to account for the 
mass budgets of the regular satellites of 
Jupiter and Saturn. This process can result 
in rock/ice fractionation as long as the make up
of the population of disk crossers is non-homogeneous, 
thus offering a natural explanation 
for the marked compositional differences between
outer solar nebula objects and 
those that accreted in the subnebulae of the
giant planets. For
a given size, icy objects are easier to capture
and to ablate, likely resulting in an overall
enrichment of ice in the subnebula. 
Furthermore, capture and
ablation of rocky fragments become inefficient far 
from the planet for two reasons: the gas surface
density of the subnebula is taken to drop outside the
centrifugal radius, and the velocity of interlopers
decreases with distance from the planet. Thus,
rocky objects crossing the outer disks of Jupiter and
Saturn never reach a temperature high enough to
ablate either due to melting or vaporization, and capture
is also greatly diminished there. In contrast, icy
objects crossing the outer disks of each planet ablate
due to the melting and vaporization of water-ice.
Consequently, our model leads to
an enhancement of the ice content of Iapetus, 
and to a lesser degree those of Titan,
Callisto and Ganymede, and accounts for the 
(non-stochastic) compositions of these large, low-porosity
outer regular satellites of Jupiter and Saturn.
For this to work, the primordial population
of planetesimals in the Jupiter-Saturn region
must be {\it partially} differentiated, so that
the ensuing collisional cascade produces
an icy population of
$\gtrsim 1$ m size fragments to be ablated during
subnebula crossing.
We argue this is likely because
the first generation 
of solar nebula $\sim 10$ km
planetesimals in the Jupiter-Saturn 
region incorporated significant 
quantities of $^{26}$Al. This is the first 
study successfully to provide 
a direct connection between nebula planetesimals
and subnebulae mixtures with 
quantifiable and observable 
consequences for the bulk properties of the regular 
satellites of Jupiter
and Saturn, and the {\it only} explanation 
presently available 
for Iapetus' low density and 
ice-rich composition.

\end{titlepage}

\section{Formation of the Regular Satellites of Giant
Planets}

The regular satellites orbit in the prograde direction, 
meaning the same sense of rotation 
as the spin of the primary, and lie close to the Laplace 
surface, with the exception of Saturn's Iapetus  
(whose orbit is nearly circular with an eccentricity of
$e = 0.028$ but inclined $i = 7.5^o$ with
respect to the Laplace surface). 
Broadly speaking, the interpretation of these 
observations is straightforward: unlike the 
irregular satellites, 
which tend to have inclined and distant orbits 
and represent a population of captured objects, 
the regular satellites form in 
circumplanetary disks of gas and solids. 
Notably, the regular satellite systems of Jupiter and
Saturn share a number of similarities, including 
the mass ratio of the largest satellites to the primary, 
the specific angular momentum, 
and bulk compositions.
Yet, the differences are also striking: a trend of 
decreasing density with 
radial distance is apparent for the Jovian but not 
the Kronian regular satellites; and 
the mass ratio between Ganymede and the 
other Galilean satellites 
is not nearly as extreme as that between Titan and 
its neighbors. Indeed,
Titan's isolation, the compositional diversity of 
Saturn's inner moons as well as its 
ring system may attest to titanic clashes between 
the well-behaved residents
proper and unruly foreign hordes -- 
interlopers from the
outer solar system wreaking-havoc upon 
the planet's orderly retinue.
In its remote outpost, Iapetus, a survivor of the onslaught, 
icily bears witness to
that early violent epoch, as its battle-scarred surface records.
Iapetus' extreme albedo contrast, large separation from Saturn and other 
Saturnian moons, its low eccentricity yet significant inclination, 
its frozen-in shape and equatorial ridge, 
its ancient, cratered surface, and icy composition all combine to make
it a Rosetta moon.
It is to this mysterious, two-faced sentinel of the outer solar
system that we primarily turn our attention.

Observational evidence indicates
that the largest KBOs are of different composition than
the regular satellites of the giant planets. Triton ($\rho = 2.061
\pm 0.007$ g cm$^{-3}$; Person et al. 2006), Eris
($\rho = 2.3 \pm 0.3$ g cm$^{3}$; Brown and Schaller 2007), and 
Pluto-Charon ($\rho = 1.94 \pm 0.09$ g cm$^{-3}$; Buie et al. 2006)
have densities that imply a rock/water-ice ratio of approximately
$70/30$ by mass. These high densities have long been
interpreted to result from accretion in the outer solar system
given the presence of abundant, oxygen-sequestering uncondensed CO 
(Prinn and Fegley 1981; McKinnon et al. 1997).
In contrast, Ganymede ($\rho = 1.94$ g cm$^{-3}$) and
Callisto ($\rho = 1.83$ g cm$^{-3}$) each consist of $\sim 50/50$ 
rock/water-ice by mass, with Ganymede being slightly more rock-rich
(Sohl et al. 2002). Similarly, Titan ($\rho = 1.88$ g cm$^{-3}$) is
also half-rock/half-ice by mass. The simplest explanation of these
observations is that the subnebulae of the giant planet were enriched
in water-ice compared to the outer solar nebula, although this interpretation
is partly clouded by uncertainties in our present understanding of
solar abundances (e.g., Wong et al. 2008).

Further insight can be sought from other regular satellites.
The inner satellites of Jupiter, Io and Europa, may have lost volatiles
either due to the temperature gradient in the subnebula (Pollack
et al. 1976; Lunine and Stevenson 1982), collisional processes
involving differentiated objects,
and/or the Laplace resonance\footnote{Amalthea may have incorporated
water-ice (Anderson et al. 2005), but it is unclear how much. Amalthea 
may be too small and close to the planet to provide
a useful constraint on satellite formation theories.}. 
Switching over to Saturn, the observed densities 
of the medium-sized regular satellites are not compatible 
with solar composition
(Wong et al. 2008). It has long been argued that these observations
are consistent with accretion in a reducing subnebula with
efficient conversion of CO to CH$_4$ 
(as opposed to inhibited conversion in the solar nebula; 
Prinn and Fegley 1981). However, satellite formation models
do not provide an environment that can support 
such chemistry (Estrada et al. 2009).
Also, close-in to Saturn collisional processes
and/or resonances (at least in the case of Enceladus) appear to
have resulted in a stochastic component\footnote{See Mosqueira
and Estrada 2003b for an alternative explanation for the ice-rich
composition of close-in, mid-size Saturnian regular satellites.}. 

Yet, Iapetus would not be affected by these processes,
and it is large enough to provide a sample of the composition of
the outer disk of Saturn.
Iapetus is different from other regular
satellites largely by virtue of its distance 
from the planet: it may have survived 
the early bombardment because impact speeds 
at its location are significantly slower 
than those closer in to Saturn, where the gravitational
well leads to hypervelocity impacts; 
its frozen-in shape and unique 
ridge are likely associated with its location far
from Saturn (Castillo-Rogez et al. 2007);  
its significant inclination is likely connected with
the deviation of the Laplace plane away 
from the planet's equator plane at its
location (Ward 1981 but note that this
scenario requires a subnebula dispersal
time of $\lesssim 10^3$ yr; Tremaine et al. 2009); 
its large albedo contrast is likely ultimately
due to the deposition of exogenous material on the 
surface of this tidally-locked regular satellite 
(Burns et al. 2009; Buratti et al. 2009); 
and, as we show here,
its composition too is 
tied to its location in the Kronian subnebula.
These properties taken together with the
difficulty inherent in the capture of a 
such a large body into a
low-eccentricity orbit provide a compelling 
case for {\it in situ} accretion 
in a circumplanetary disk of gas and 
dust\footnote{Elsewhere 
(Mosqueira and Estrada 2005), we 
discuss an intriguing collisional-scattering scenario for the
origin of Iapetus; however, we favor the present model.}.

To establish a connection to the
outer Kronian subnebula we must first constrain Iapetus' porosity.
Given its mean radius and mean
density ($r = 736.0 \pm 2$ km and $\rho = 1.081 \pm 18$ g cm$^{-3}$;
Thomas et al. 2006) the pressure at
the core ($P = (2 \pi/3) \rho^2 Gr^2 \simeq 100$ MPa) is high enough
that porosity should be small there (though it could be larger 
in the outer layers of the satellite). Experimental
results indicate a reduction in porosity in cold, granular water
ice over a pressure range of $\sim 1-150$ MPa, with porosities
of the order of $\sim 10 \%$ for $\sim 100$ MPa pressures (Durham
et al. 2005). Heating, annealing and creep-driven 
water-ice flow can result in pore collapse and decrease the 
porosity even more. Models in which Iapetus de-spins due
to the presence of short-lived radioactive isotopes (SLRI),
leading to the formation of a ridge, imply warm-ice and 
low porosities (Castillo-Rogez et al. 2007; Melosh and Nimmo 2009).
Although the rheology of icy satellites is poorly understood,
the interpretation of Iapetus' fossil shape as ``frozen-in''
due to lithospheric hardening as the satellite was in the
process of despinning is persuasive, and at odds with with a 
high-porosity object. Note that weak radiogenic heating 
due to long-lived radioactive isotopes (LLRI) alone would be enough
to collapse the pores except in a surface layer tens of kilometers thick
(Castillo-Rogez et al. 2007). An Iapetus' model with a porosity
of $\sim 0.3$ in a clean-ice upper $\sim 50$ km (roughly the surface layer thickness
of the Castillo-Rogez et al. (2007) LLRI model; cf their Fig. 6) of the satellite would
result in a rock ratio of $\sim 30 \%$ by mass, which is still
rock-depleted compared to Ganymede, Titan and Callisto.
Furthermore, even if it were possible to
despin a cold Iapetus, the presence of the ridge is likely to 
indicate ice mobility early on. Thus, we argue
that Iapetus is an icy satellite with rock/water-ice fraction
of $\sim 20 \%$ by mass, making its composition
incompatible with solar mixtures (Wong et al. 2008).

We stress that Iapetus' low
density relative to Titan, Ganymede and Callisto 
{\it cannot} be explained by a snowline
argument since all four satellites are taken to accrete
outside the snowline (Pollack 1976; Lunine and Stevenson 1982).
The issue here is the depletion of rock, as 
opposed to the presence of volatiles, in Iapetus.
In this regard, it is instructive to compare Iapetus to irregular Phoebe. 
The Cassini flyby of Phoebe yielded a density of $1.630 \pm
0.046$ g cm$^{-3}$ (Jacobson et al. 2006).  If we allow for
a moderate porosity, which is reasonable given
its $106.60 \pm 1.00$ km size, this density 
corresponds to a rock/water-ice
fraction similar to that of large KBOs.
The contrast between Iapetus and Phoebe reinforces the
interpretation of Phoebe as a captured moon from the
outer solar nebula (Johnson and Lunine 2005).
Indeed, the rock/water-ice fraction for Phoebe may
be larger than those of Ganymede, Callisto and Titan.
Allowing for significant porosity in the case
of Phoebe would accentuate 
the compositional contrast with Iapetus
and other regular satellites.

In the nucleated instability model of planet 
formation (e.g., Bodenheimer and Pollack 1986; Pollack et al. 1996) 
a core must first form by accretion of planetesimals. 
In this mode of planet formation 
most of the mass of solids resides
in planetesimals of size $\sim 10$ km 
(Wetherill and Stewart 1993).
Planetesimals are also needed
to explain the observations of the Oort cloud and the scattered
belt (e.g., Charnoz and Morbidelli 2007), to power planet migration
in the Nice model (e.g., Morbidelli et al. 2008), and to explain the
volatile enhancement in giant planet atmospheres by
planetesimal trapping and delivery (e.g., Owen and Encrenaz 2003).
Here we seek to provide a direct physical link between 
planetesimals in the solar nebula 
and the circumplanetary disks of giant planets, and to 
account for the source of the solids 
that ultimately led to the formation of the regular satellites 
of the giant planets. Our aim is to study mass delivery by ablation of 
planetesimal fragments crossing the circumplanetary disk
(to be followed by re-condensation and satellite accretion). 
In particular, ablation can result in fractionation, 
and account for the observed density of Iapetus 
provided that this satellite formed {\it in situ} 
(Mosqueira and Estrada, 2005).
Although we focus on Iapetus, our model applies 
to the origin of the regular 
satellites in general\footnote{For instance, the puzzling observation that primordial
$^{36}$Ar is enhanced in the atmosphere of Jupiter (Atreya et al. 1999), 
but only present in trace amounts in the atmosphere of Titan, as reported
by the GCMS instrument aboard Cassini (Niemann et al. 2005), may
fit within a framework of planetesimal ablation followed by subnebula
re-condensation of some volatiles but not others,
but we leave this subject for further work.}. The model we present here fits
with the gaseous SEMM satellite formation model 
(Mosqueira and Estrada 2003a,b; hereafter MEa,b; see also Estrada et al. 2009 and
Mosqueira et al. 2009). This framework is consistent with the core accretion
planet formation model in which Jupiter and Saturn form
$\sim 3-5$ Myr after CAIs (Hubickyj et al. 2005; Dodson-Robinson et al. 2008).
Note that in alternative 
{\it gas-poor} or {\it gas-starved} satellite formation models planetesimals 
do not traverse enough 
subnebula gas to ablate significantly. Thus, mass delivery by ablation is not available to
these models whether they consider planetesimals as a source of solids (Estrada and Mosqueira
2006) or not (Canup and Ward 2002).

For rock/water-ice fractionation to take place, and
our Iapetus' model to obtain, the first generation
of planetesimals in the Jupiter-Saturn region must
incorporate significant quantities
of $^{26}$Al, which depends on their time of formation.
The closest analog to the planetesimals forming 
in the Jupiter-Saturn region is the asteroid belt. The ages
of asteroids indicate that they formed 
with a spread of $1-3$ Myr after Ca-Al-rich inclusions 
(CAIs) (Scott and Krot 2005). 
The asteroid belt may have been cleared largely by scattering due to 
planetary embryos and a migrating Jupiter
(Bottke et al. 2005). Yet, 
the fossil evidence provided by asteroids may not
be directly linked to the first generation of 
planetesimals. (It is worth noting that
the parent bodies of iron meteorites may have
formed $1-2$ Myr earlier than those of the ordinary
chondrites; Baker et al. 2005.)
Instead, main belt asteroids are 
the product of a few My of accretion
before most of the mass in the belt was depleted 
(Weidenschilling 2009). Indeed, as runaway growth
proceeds to completion most small bodies would be 
ground down by embryos. The degree of collisional
evolution in asteroidal objects depends on the size of 
the planetesimals, which may
be used to provide a constraint of 
$\sim 10 - 100$ km on primordial planetesimal 
sizes (Weidenschilling 2009). Thus, neither asteroids nor meteorites 
provide a direct sample of this population. Nevertheless,
iron meteorites may have come 
from $\sim 20-200$ km differentiated parent 
bodies that underwent 
catastrophic collisions (Chabot and Haack 2006).

The timescale for the formation of planetesimals depends
on the nebular environment early on. Assuming 
efficient sticking, the timescale for the formation of
planetesimals is very short indeed $\tau \sim 2000$
$(a/1\, \rm{AU})^{3/2}$ years, where $a$ is the radial distance from
the Sun (Weidenschilling 2000). Collisional growth appears
to be efficient for small particle sizes and impact speeds 
(see review by Dominik et al. 2007); yet, growth past the
decoupling size $\sim 1$ m in the presence of
turbulence is questionable 
(Mosqueira 2004; Cuzzi and Weideschilling 2006). 
The problem is especially
severe in light of the short clearing times of $\sim 1$ m size objects.
It can be argued that in the presence of turbulence planetesimals
can bypass this size range entirely (Johansen et al. 2007;
Cuzzi et al. 2008), however 
there is no agreement regarding the
specific mechanism to accomplish this leap -- the simulations
of Johansen et al. (2007) {\it assume} a particle size that is
already partially decoupled from the gas,
whereas the simulations of
Cuzzi et al. (2008) do not yet show that a large planetesimal
$\gtrsim 10$ km can actually form 
(even sporadically) from chondrule-like
$\sim 1$ mm sized particles. While a recent study concludes
that asteroids are born big (Morbidelli et al. 2009), given
uncertainties in the strength of planetesimals and the
role of collisional debris, the asteroid size distribution, and
the timing of the excitation caused by the migration of Jupiter,
planetesimal sizes $10-100$ km appear consistent with
the data. In this regard, the possibility should be
explored that smaller planetesimals are disrupted by planetary
embryos prior to the final stages of asteroid accretion.

The alternative that quiescent regions can help to form
planetesimals has been considered in a number of publications
(Kretke and Lin 2007; Brauer et al. 2008; Lyra et al. 2009). 
The stumbling
block here may be an embarrassment of riches, i.e., in a
laminar disk it is {\it too easy} to make planetesimals. Consequently,
dust might coagulate into larger objects too fast to
explain the observations of protoplanetary
disks indicating the presence of small $\lesssim 3$ $\mu$m grains
(e.g., Dullemond and Dominik 2005; 
Cuzzi et al. 2008; Birnstiel et al. 2009), the evidence for {\it delayed}
($\sim 1$ Myr) grain growth based on observations of 
SEDs of young disks (Beckwith
et al. 2000; Currie et al. 2009), or the
lack of pervasive differentiation of main belt asteroids (Cuzzi et al. 2008).
However, dust may be replenished by weak inflow
onto the protoplanetary disk (Dominik and Dullemond 2008),
by viscous mixing from outside the ``dead-zone''
(Birnstiel et al. 2008), or by collisional
grinding; in addition, main belt 
asteroids do not constitute a fossil record of the first generation of 
planetesimals. The degree of activity in the ``dead-zone'' 
remains to be clarified (Turner et al. 2007; Bai and Goodman 2009), 
but there is general agreement in the literature 
that disk quiescence facilitates 
planetesimal formation (e.g., Youdin and Shu 2002; Cuzzi and
Weidenschilling 2006). Since the protoplanetary disk may
remain turbulent and variable during its initial stages (the
``FU-Ori'' epoch) lasting
a few $10^5$ years (Bell et al. 2000), 
we take $\lesssim 10^6$ years to be the planetesimal 
formation timescale, implying
that these planetary building blocks 
incorporate significant quantities of short-lived
radioactive elements.

In Sec. 2 we discuss the framework and model parameters
for the computation of the mass delivery to the circumplanetary
disk. In Sec. 3 we describe the compositional constraints
that our model seeks to address, and the physical processes used to
that end. We briefly discuss the thermal evolution of planetesimals
in the Jupiter-Saturn region due to the presence of SLRIs in Sec. 4. 
We close with our conclusions in Sec. 5.

\section{Model parameters}

Following giant planet formation 
planetesimals in its feeding zone undergo collisional grinding. 
In the Jupiter-Saturn region the collisional 
timescale for kilometer-sized objects is similar 
to the ejection timescale $\sim 10^4$ yrs, so that a fraction of the 
mass of solids will be fragmented into objects 
smaller than 1 km (Stern and Weissman 2001; Charnoz and Morbidelli 2003).
The collisional cascade facilitates planetesimal delivery to the circumplanetary
disk because smaller planetesimals are easier to capture.
An estimate of the size that will be fragmented may be obtained
by equating the ejection time $\tau_{eject} \sim 0.1 \,\Omega^{-1}
(M_\odot/M_P)^2$, where $\Omega$ is the angular velocity, $M_P$ is the
planet's mass and $M_\odot$ is the Sun's mass, to the collision time
$\tau_s \sim \rho_s r/\Omega \sigma_s$, where
$\rho_s$ is the density of the solids, $r$ is the particle size and $\sigma_s$
is the surface density of the solids. This yields particles as large as
$10$ km at Jupiter (and larger for Saturn)
for surface densities a few times the minimum mass (Hayashi 1981).

Switching to the subnebula, in the SEMM model (MEa,b) one expects a factor
of $\sim 10$ enhancement in solids over cosmic mixtures, resulting
in a gas surface density of $\sim 10^{4}$ g cm$^{-2}$ (see Sec. 2.2), which
is consistent with and quantitatively constrained 
by the gap-opening condition for Ganymede and
Titan in a inviscid disk with aspect ratio $\sim 0.1$ (Rafikov 2002; MEb),
and the Type I migration of full-sized satellites in such
a disk (MEb; Bate \etal 2003). If we use a gas surface density of 
$10^4-10^5$ g cm$^{-2}$ for the Jovian
and Saturnian subnebulae,
then a planetesimal of density $\sim 1$ g cm$^{-2}$ will encounter a gas
column equal to its mass if its radius is in the range $0.1 - 1$ km, resulting
in ablation and delivery to the circumplanetary disk.

Our aim here is to provide a
link via ablation of planetesimal fragments in the
Jupiter-Saturn dynamically active region to the
circumplanetary disks of these two planets, seeking
to account for the mass and angular momentum
budgets and compositions of the regular satellites. 
Tying solar nebula planetesimals to subnebula satellitesimals
involves three different aspects: first, characterizing the properties of the 
first generation of planetesimals in terms 
of sizes and degree of heterogeneity; 
second, quantifying the collisional evolution 
of the planetesimal swarm following giant planet formation; 
and third, delivering planetesimal fragments 
to the circumplanetary disks of Jupiter and Saturn.

\subsection{Planetesimal fragmentation and the initial size distribution}

The strength of planetesimals is an unknown quantity.
A strain-rate scaling law by Housen \etal (1991) suggests that
kilometer-sized objects are the weakest.
Davis and Farinella (1997) find a constant
$S = 3 \times 10^6$ erg cm$^{-3}$, where $S$ is the impact
strength ({\it i.e.}, the energy per unit volume for
shattering fifty percent by mass of the parent body) for crushed icy bodies,
in agreement with Ryan \etal (1999). Benz and
Asphaug (1999) use a 3D Hydrocode to simulate $3$ km/s
impacts and find that $100$ m objects may be weakest, with kilometer
sized planetesimals strengthened by gravity. How much planetesimal
mass fragments following the collisional cascade and is delivered
depends on the planetesimal strength (Charnoz and Morbidelli 2003).

The initial size distribution is
also unknown. If run-away growth takes place the resulting
size distribution may be steep. Wetherill and Stewart
(1993) find a bimodal size distribution with a fragmentation
tail of objects less than $1$ km with power law exponent
$q \sim 3.5$ and size distribution for objects larger
than that with $q \sim 5.5$, where the differential
size distribution ($dN/dr \propto r^{-q}$) approaches
a value of $q \sim 3.5$ in the case of a size independent
fragmentation model (e.g., Dohnanyi 1969).
In his coagulation simulations Weidenschilling (1997) finds
that most of the mass resides in objects $\sim 10$ km in a
timescale of $10^5 - 10^6$ years in the outer nebula. 
Most studies of planet 
formation start with most of the mass of solids in objects
of size $\sim 10$ km, as suggested by numerical simulations
(e.g., Kenyon and Luu 1999). 

Unless otherwise stated, we consider values of
$q = 3.5$. We set the lower size cut-off at $1$ m. The
reason is that smaller objects are protected from further collisional
grinding by gas drag, so that below this size collisions are
once again accretive as long as the nebula turbulence is
weak. We vary the upper size cut-off between
$10-100$ km {\it in lieu} of more detailed collisional fragmentation
simulations, which we leave for further work.
For simplicity, we also assume a fully fractionated population of
icy and rocky (which includes iron) planetesimal fragments,
each of which has the same particle size spectrum (see Sec. 4). 

\subsection{Disk properties}

Gas flowing into its Hill sphere forms a disk
around the protoplanet. The timescale for Jupiter and
Saturn to clear a gap in between $\sim 10^4$ years sets
the time for the end of gas accretion. The timescale
for envelope collapse is given by the
Kelvin-Helmholtz time of $10^4-10^5$ years (Hubickyj \etal 2005),
following which a circumplanetary disk forms. {\it Prior} to gap-opening
the giant planet accretes gas with semi-major axis
originating from $\sim R_H$ of its location, 
where $R_H = a(M_P/3M_\odot)^{1/3}$ is the planet's Hill radius,
centrifugal balance
yields a characteristic disk size of $\sim R_H/50$ (Stevenson \etal 1986; MEa). 
For Jupiter and Saturn these radii are located close to the positions of
Ganymede and Titan. {\it After}
gap-opening accretion continues through the planetary Lagrange
points, and the estimated characteristic disk size is
larger $\sim R_H/5$ (MEa; Estrada \etal 2009; Ayliffe and Bate 2009), which
we take to be the radial extent of the circumplanetary disk. This radial size 
corresponds to $\sim 150$ R$_J$, and $\sim 200$ R$_S$, where $R_J$ and 
$R_S$ are the radii of Jupiter and Saturn, respectively.

MEa divide the circumplanetary disk into inner and outer regions. 
For Jupiter, we compute the solids-enhanced minimum mass (SEMM) 
gas densities in the
inner and outer disks based on the
solid mass required to form Io, Europa (both re-constituted for missing
volatiles) and Ganymede in the
inner disk, and Callisto in the outer disk.
Inside of the centrifugal radius
the surface gas density is $\Sigma \sim 
10^4-10^5$ g cm$^{-2}$, which corresponds to pressures $\sim 0.1$ bar.
Outside the centrifugal radius $R_c \sim 30 \,{\rm{R}}_J$,
the gas surface densities are in the range $10^2-10^3$ g cm$^{-2}$.
We apply the same procedure to the circumplanetary disk
of Saturn by employing the masses of Titan and
Iapetus to set the inner and outer disk masses, respectively. 
The transition region has a width $\gtrsim 2H_c$, where
$H_c$ is the subnebula scale-height at the centrifugal radius.
This choice ensures that the
gradient in gas density is not so steep as to lead to a Rayleigh-Taylor
instability (e.g., Lin and Papaloizou 1993).

The subnebula gas surface density profile is taken to be

\begin{equation}
\Sigma(R) =
\begin{cases}
   \Sigma_{a} (R_{a}/R),     &\,R < R_a;\\
   \Sigma_{a} (R_{a}/R) - \frac{\Sigma_{a} (R_{a}/R)-\Sigma_{b} (R_{b}/R)^2}{e^\frac{R_a+R_b-R}{2n}+1},             &\,R_a < R < R_b;\\
   \Sigma_{b} (R_{b}/R)^2,   &\,R > R_b.\\
\end{cases}
\end{equation}

Once a disk forms, the cooling timescale depends on the opacity.
As particles grow, the opacity of the nebula decreases, allowing
the gas to cool and accretion of ice-rich satellites to take place.
We connect the disk's temperature profile to the planetary luminosity
at the tail end of giant planet formation
(Hubickyj \etal 2005). We use a heuristic temperature profile of the form
$T = 3600\,{\rm{R}}_J/R$ for Jupiter (e.g., Lunine and Stevenson
1982) and $T = 2000 \,{\rm{R}}_S/R$ for Saturn. The outer disks of
Jupiter and Saturn have roughly constant temperatures
in the range of $70 - 130$ K for Jupiter and
$40 - 90$ K for Saturn, depending on solar nebula parameters.

\section{Regular Satellite Mass and Composition Constraints}

The compositional gradient of the Galilean 
satellites may provide a probe to the environment 
in which they formed (Estrada et al. 2009 
and references therein). 
Here we focus on the large, 
outer regular satellites of each satellite system: 
Ganymede and Callisto in the case of Jupiter; 
and Titan and Iapetus for Saturn. Planetesimal 
break-up in tandem with delivery via ablation of 
planetesimal fragments crossing the subdisk 
provides a framework for understanding the mass 
budget and compositions of regular satellites 
compared to that of solar nebula planetesimals. 

\subsection{Thermal Ablation of Disk Crossers}

Thermal ablation occurs because friction 
of the fast moving heliocentric interloper 
as it crosses the circumplanetary gas 
disk heats up the body. At low gas densities 
impactors lose mass and energy through ablation. 
For large kilometer-sized bolides mechanical 
destruction rather than thermal ablation 
may dictate the fate of the object 
(e.g., Zahnle and Mac Low 1994). 
Ablation of planetesimal fragments $< 1$ km 
may deliver the bulk of the solids 
needed to form the satellites. 
For objects in that size range, 
the heat transfer coefficient 
$C_H \sim 0.1$ may be reasonably obtained 
from observations of terrestrial meteorites 
(Bronshten 1983). The rate at which energy 
is transferred to the planetesimal per unit area
is given by $E \sim 0.5 C_H \rho v^3$, where $\rho$ is 
the gas density, $v$ is the speed of the 
planetesimal through the gas, and some 
planetesimal flattening (which increases its cross section) 
takes place (e.g., Zahnle 1992; Chyba \etal 1993). 
Ablation thus results in 
delivery of material to the circumplanetary disk.
Gas drag can also result in capture of material. 

Ignoring conduction into the interior and ablation, we can obtain
the surface temperature $T_{surf}$ by balancing this heating
and radiative cooling $\epsilon \sigma_{SB} T_{surf}^4$,

\begin{equation}
T_{surf} = \frac{1}{\epsilon^{1/4}} \left( \frac{f C_H}{8 \sigma_{SB}} \rho v^3 + T_0^4 \right)^{1/4},
\end{equation}

\noindent
where $\epsilon=0.5$ is the emissivity, $T_0$ is the 
subnebula background temperature,
$\sigma_{SB}$ is 
the Stefan-Boltzmann constant,
$v$ is the speed of the planetesimal through the gas disk, and $f$ is
a parameter that measures the degree of flattening. It is important to
keep in mind that this quantity reflects the energy regime not the
actual surface temperature unless it is low
enough that ablation can be ignored.
Using parameters appropriate at the location of Iapetus
$v \sim 5 \times 10^5$ cm $s^{-1}$ (a cold population at infinity), $f = 4$ and 
$\rho = \Sigma/ 2H \sim 7 \times 10^{-10}$ g cm$^{-3}$, where
$\Sigma$ is the gas surface density and $H=c/\Omega$ is the scale-height,
$c$ the sound speed and $\Omega = \sqrt{GM_P/R^3}$ is the angular
velocity in the circumplanetary disk, we obtain a surface temperature of
$T_{surf} \sim 900 ^o$ K, which is
sufficient to melt and vaporize icy
objects but not rocky objects. A value of $f=1$, which is
suitable to the outer disk due to the low ram pressures there,
yields a temperature of $\sim 600 ^o$ K. In contrast, at
Titan ($v \sim 8 \times 10^5$ cm $s^{-1}$; $\rho \sim 3 \times 10^{-7}$ g cm$^{-3}$)
and Callisto ($v \sim 1.1 \times 10^6$ cm $s^{-1}$; $\rho \sim 2 \times 10^{-7}$
g cm$^{-3}$) the surface
temperature can exceed $4000 ^o$ K (for $f =4$), which is
enough to melt and vaporize rocky objects as well.
Hence, the temperature of objects crossing the disk places
Iapetus in a separate regime from other regular
satellites, i.e., at its location in the disk rocky objects do not
reach temperatures high enough to ablate.


We can estimate the change in radius of an icy planetesimal
due to either melting or vaporization as it crosses
the gas disk in a time $\Delta t \approx 2 H/v$ (ignoring gas drag
for now) (Podolak \etal 1988)

\begin{equation}
\left(\frac{dr}{dt}\right)_{vap,melt} \approx 
- \frac{fC_H\rho v^3}{8\rho_pE_{0c,m}}
\left\{1 - \frac{8\sigma_{SB}T_{0}^4}{fC_H\rho v^3} \left[\epsilon (T_{c,m}/T_0)^4 - 1\right]\right\},
\end{equation}

\noindent
where $\rho_p$ is the planetesimal density, $T_{c} = 648 ^o$ K is the
critical temperature for vaporization, and $T_m = 273 ^o$ K is the
melting temperature. $E_{0m} = 6.0 \times 10^{9}$ erg g$^{-1}$ is 
the energy required to raise the temperature of 1 g of ice from the
background temperature of the nebula $\sim 100 ^o$ K (where it originated)
to $T_m$ plus the latent heat of melting $H_f$, 
and $E_{0c} = 2.8 \times 10^{10}$ erg g$^{-1}$
is the energy needed to raise the temperature of $1$ g of icy
material for $100 ^o$ K to $T_c$ (so that vaporization
becomes energy limited) plus the latent heat of vaporization $H_v$. 
We find that the amount of ice melted or vaporized at Iapetus ($\Delta t \sim
3\times 10^5$ s) is in the order of meters (for $f=1$), whereas at Titan and Callisto ($\Delta t \sim
4\times 10^4$ s) it is about a kilometer (for $f=4$). 
It should be noted that at Iapetus ($\Sigma \sim 100$ g cm$^{-2}$)
meter-sized objects (with density $\sim 1$ g cm$^{-3}$) traverse a column
of gas equal to their mass, whereas at Titan and Callisto
($\Sigma \sim 10^4$ g cm$^{-2}$) the
same is true for kilometer-sized objects.
Hence, ablation of planetesimal fragments
may provide the solids needed for
the formation of the regular satellites provided that a significant fraction
of the planetesimal mass resides in fragments with sizes
in the kilometer to meter size range.

One can ask whether a particle will ablate versus breakup
when crossing the disk. At Titan,
the dynamic pressure is $P_{dyn} = \rho v^2/2 \sim 10^5$
dynes cm$^{-2}$. At Iapetus, we find $P_{dyn} \sim 100$ dynes cm$^{-2}$.
Therefore, if a meter-sized or larger particle with strength
$S \sim 10^6$ dynes cm$^{-2}$ crosses the disk icy
material is left behind by ablation (at Titan and Iapetus) or by gas drag
capture (at Titan), but probably not breakup (at least at Iapetus). 
It is also important to consider the sensitivity of our results to the
flattening parameter $f$. In this regard, it should be noted
that the overall dependence of ablation on $f$ is not too strong, and that
at the location of Iapetus the ram pressure is low enough that
$f = 1$ is a good approximation. 

To find out the mass budget and composition of the subnebula resulting
from the ablation of an unmixed population of rocky and
icy disk crossers, we have to tackle the thorny issue
of whether the rate of ablation in Eq. 3 
is dominated by melting or vaporization. It is probably fair
to say that most models of the ablation process
focus on vaporization (e.g., Moses 1992). However, at lower
temperatures the rate of ablation may be affected by the viscosity
of the melt and other factors (see Podolak et al. 1988 for a discussion).
Yet, a detailed model incorporating the full complexity of the
problem for a broad range of parameter space
is premature in that it can obscure our conclusions. 
In fact, as we show below, our main results are robust
and not dependent on the specifics of the ablation mechanism.
For this reason, we leave development of a more detailed
ablation model for later work, and we employ the classic
approach of Bronshten with minor modifications.
We calculate the amount of material ablated by solving coupled equations of
motion and ablation (Bronshten 1983) for a planetesimal of mass $m$ which is assumed for
simplicity to cross the circumplanetary disk perpendicular to the disk plane:

\begin{equation}
m\frac{dv}{dt} = -\frac{1}{2}C_DA(m)\rho(X,Z)v^2 - m \Omega^2(X) z;
\label{equ:mdvdt}
\end{equation}

\begin{equation}
\begin{split}
\frac{dm}{dt} = -\frac{C_H}{2E_{0c,m}}A(m)\left[\rho(X,Z)v^3 - \frac{8 \sigma_{SB}}{C_H} (\epsilon T_{c,m}^4 - T_0^4)\right]\,\,\, {\rm{for}}\,\,\, T_{surf} > T_{c,m}; \\
\frac{dm}{dt} = 0\,\,\, {\rm{for}}\,\,\, T_{surf} \leq T_{c,m},  
\end{split}
\label{equ:dmdt}
\end{equation}

\noindent
where $T_{surf}$ is obtained from Eq. 2 with $f = 1$, 
$C_D = 0.44$ is the drag coefficient, $A$ is the planetesimal
cross-section, $X = R/R_P$ is the distance from the planet in
units of planetary radii $R_P$, $Z = z/\sqrt{2}H$,
$\rho(X,Z) = \rho_0(X) e^{-Z^2}$
is the gas density of the subnebula. Here we set the flattening factor
$f = 1$ because we are particularly interested in the outer disk, where
the ram pressure is low.
Note that for $T_{surf} \leq T_{c,m}$ the ablation equation goes to
zero even though the planetesimal can still be captured
by gas drag. That is, regions of the disk such that the
disk crosser never reaches the ablation temperature $T_{c,m}$
may only be delivered to the circumplanetary disk by
gas drag capture. Fits to observational data of meteoroids in the
atmosphere yield values of $C_H \simeq 0.1$  for low
values of the effective heat of ablation 
$\sim 10^{10}$ erg g$^{-1} $ (Svetsov et al. 1995),
which leads us to use $E_{0m} = 3.0 \times 10^{10}$ erg g$^{-1}$
and an ablation temperature of $T_m = 1800 ^o$ K for rock.
For ice we consider both melting ($E_{0m} = 6.0 \times 10^9$ erg g$^{-1}$; $T_m = 273 ^o$ K)
and vaporization ($E_{0c} = 2.8 \times 10^{10}$ erg g$^{-1}$; $T_c = 648 ^o$ K)
while keeping $C_H \simeq 0.1$ constant. We stress that these parameter
choices are quite conservative and intended to test the sensitivity of
our results, as significant water-ice sublimation 
will take place for $T_{surf} < T_{c,m}$. Thus, a more realistic
treatment is likely to deliver significantly more water to the outer disk. 
 
\begin{figure}[t!]
 \resizebox{\linewidth}{!}{%
 \includegraphics{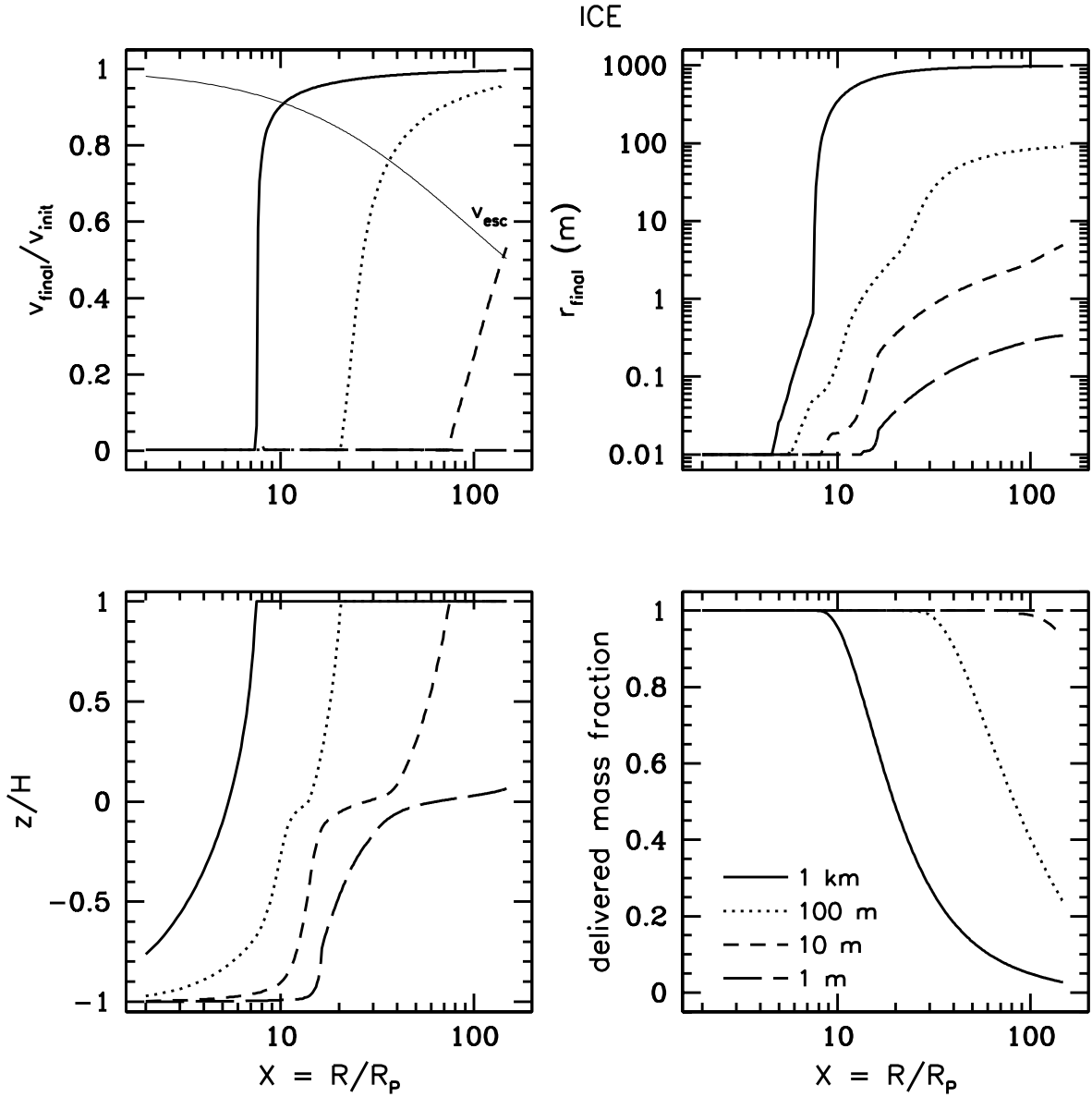}}
\caption{{\small Plot final velocity relative to initial $v_{final}/v_{init}$, final
planetesimal size $r_{final}$, distance travelled through the disk $z/H$,
and delivered mass fraction due to ablation and gas drag capture
as a function of the distance from Jupiter, $X$, for several initial planetesimal
sizes (indicated by the solid and dashed curves). The gas surface density is taken to be
that of the SEMM model, $\Sigma(X) = 3\times 10^5/X$ g cm$^{-2}$ (without a transition).
We take the initial velocity as the interloper crosses the disk to be
$v_0\approx 60/\sqrt{X}$ km s$^{-1}$. The planet's escape velocity as a function
of distance $v_{esc}$ is plotted in the top left panel.}}
\label{fig:jice}
\end{figure}

We solve Equations \ref{equ:mdvdt} and \ref{equ:dmdt} using a fourth order
Runge-Kutta scheme for coupled equations. In Fig. 1 we plot final
velocity relative to initial $v_{final}/v_{init}$, final
planetesimal size $r_{final}$, distance travelled through the disk $z/H$,
and delivered mass fraction due to ablation and gas drag capture
as a function of the distance from Jupiter, $X$, for
different particle sizes. Here 
the gas surface density corresponds to that of a SEMM
model $\Sigma(X) = 3\times 10^5/X \gcms$ (without a transition). 
These results use the more conservative ice 
vaporization values for the ablation temperature
$T_c$ and heat of ablation $E_{0c}$. If the planetesimal does not completely 
ablate, it may successfully pass
through the disk. However, planetesimals may fail to traverse the disk
because they are ablated before they do, or because of gas
drag capture. If particles are ablated down to a size of
$1$ cm, they are taken to have totally ablated. Close to
the planet even $\sim 1$ km planetesimals (bold solid lines) are completely ablated
before reaching a final velocity $v_{final} = 0$ (top left panel).
Farther out in the disk, a leftover planetesimal of size
$r_{final}$ (top right panel) emerges from the disk after
crossing a distance of $2 H$ (bottom left panel). As expected, the mass
delivered to the disk by either capture or ablation
decreases with distance from the planet (bottom right panel).

\begin{figure}[t!]
 \resizebox{\linewidth}{!}{%
 \includegraphics{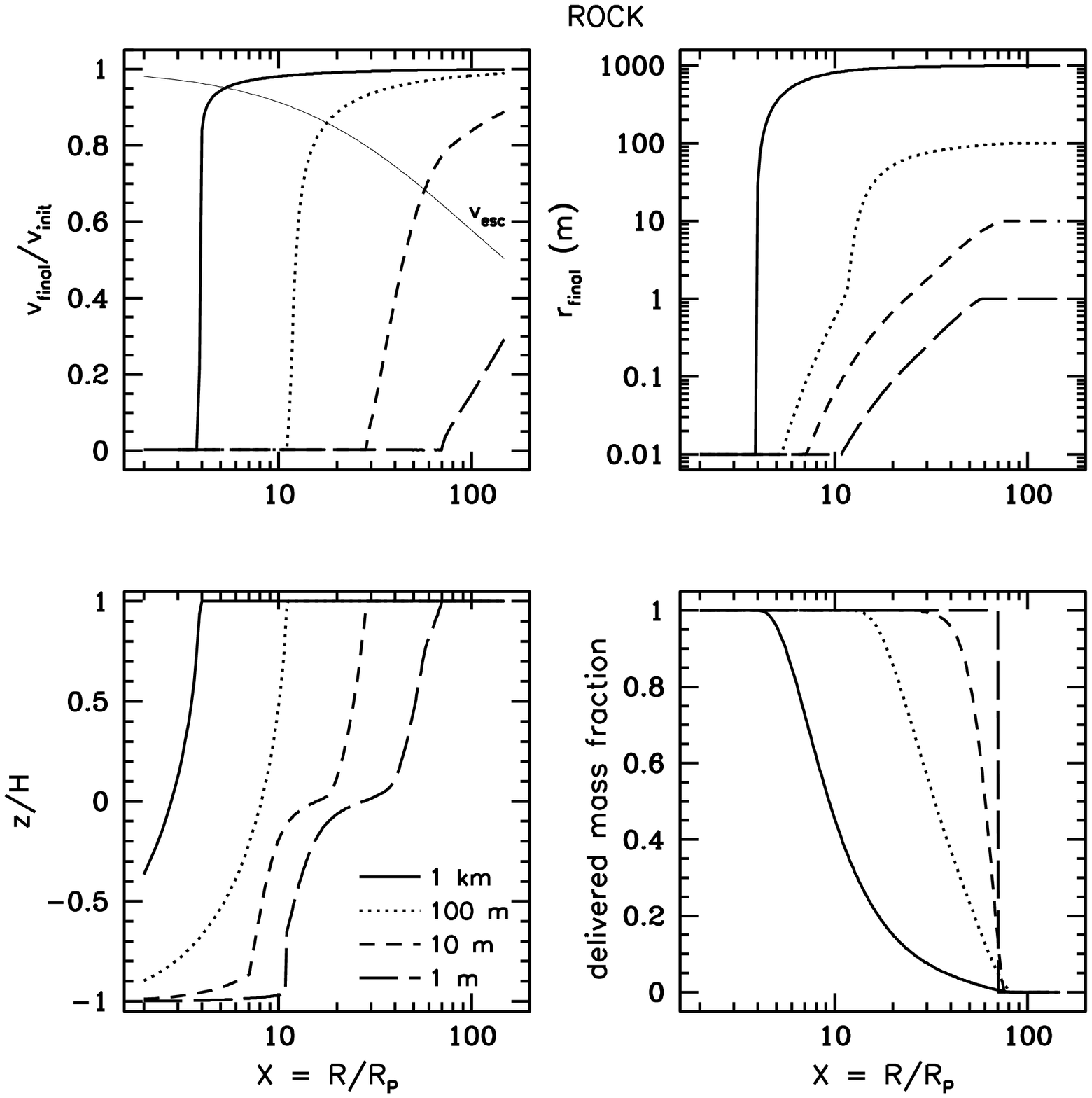}}
\caption{{\small Plot final velocity relative to initial $v_{final}/v_{init}$, final
planetesimal size $r_{final}$, distance travelled through the disk $z/H$,
and delivered mass fraction due to ablation and gas drag capture
as a function of the distance from Jupiter, $X$, for several initial planetesimal
sizes (indicated by the solid and dashed curves). The gas surface density is taken to be
that of the SEMM model, $\Sigma(X) = 3\times 10^5/X$ g cm$^{-2}$ (without a transition).
The planet's escape velocity as a function
of distance $v_{esc}$ is plotted in the top left panel.}}
\label{fig:jrock}
\end{figure}

In the top left curve of Fig. \ref{fig:jice} we also plot the escape velocity 
$v_{esc} = \sqrt{2 G M_P/(R_PX)}$
as function of distance (light solid line) for reference. It is important to
point out that here (and in the rest of this paper) 
our capture condition is defined locally. That is, we only consider that the
particle has been {\it locally} captured if $v_{final} = 0$. 
A planetesimal can emerge from the disk with a speed too
slow to escape the gravitational potential well of the planet, so that
it will be delivered either to the circumplanetary 
disk or the planet. For instance, $10$ m particles
(short dashed lines) have final speeds that are less than the escape speed
out to the edge of the disk. Therefore, these particles will be delivered
to the circumplanetary disk, but not always {\it locally}. Here
we require $v_{final} = 0$ for capture since in this case the planetesimal mass
is delivered at the location of passage. This is conservative in the sense that it 
underestimates the mass delivered to the circumplanetary disk. But it
is likely to provide a better estimate of the ice/rock ratio as a function
of position because particles with significant non-zero exit velocities can 
potentially be delivered
anywhere inside the crossing radius. We leave a more realistic treatment of multiple 
passes through the disk for later work.

In Fig. \ref{fig:jrock} we do the corresponding calculations for the case of rock.
Note that in the case of rock we employ the rock melting values
for the ablation temperature $T_m$ and heat of ablation $E_{0m}$.
For rock the mass delivered for a given particle size (bottom right
panel) drops faster than for ice despite our conservative 
choice of parameters for ice ablation. Consequently, the composition of 
disk is expected to be enriched in ice. Rock mass delivery in the outer disk can be 
dominated by capture. This can be seen by the behavior of $1$ m particles (long
dashed lines): gas drag slows down these particles sufficiently fast that almost no
ablation takes place beyond $X = 60$ so that $r_{final} = 1$ m (top right panel).
Yet gas drag is effective out to $X = 70$ despite the low surface density of 
the outer disk so that $v_{final} = 0$ (top left panel) and the full particle
mass is delivered by gas drag capture out to this location (bottom right panel).
Outside this location essentially no rock is delivered. Because
ice is still ablated and captured outside this location (Fig. \ref{fig:jice}), 
one can expect the cold outer disk to be water-ice enriched following 
re-condensation.

\begin{figure}[t!]
 \resizebox{\linewidth}{!}{%
 \includegraphics{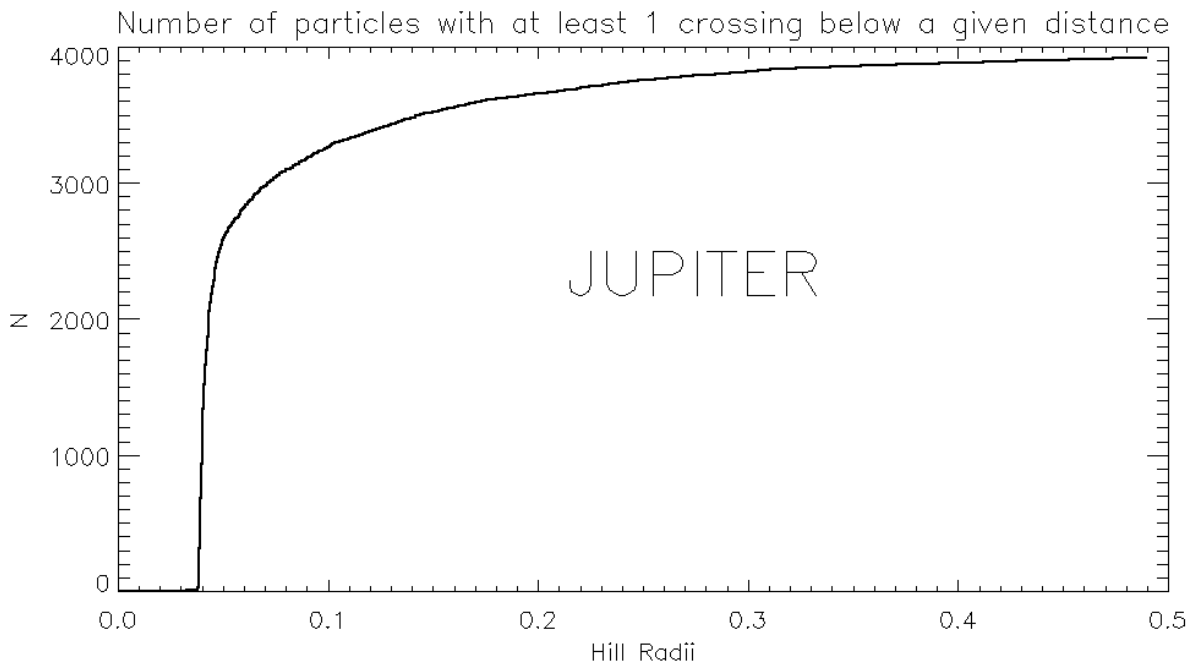}}
\caption{{\small Number of particles with at least one crossing inside a given
distance from Jupiter. Simulations involved 5000 particles with total mass of $10$
M$_\oplus$. We estimate that roughly $\sim 7$ M$_\oplus$ pass within
0.2 R$_H$.}}
\label{fig:jupcap}
\end{figure}

\begin{figure}[t!]
 \resizebox{\linewidth}{!}{%
 \includegraphics{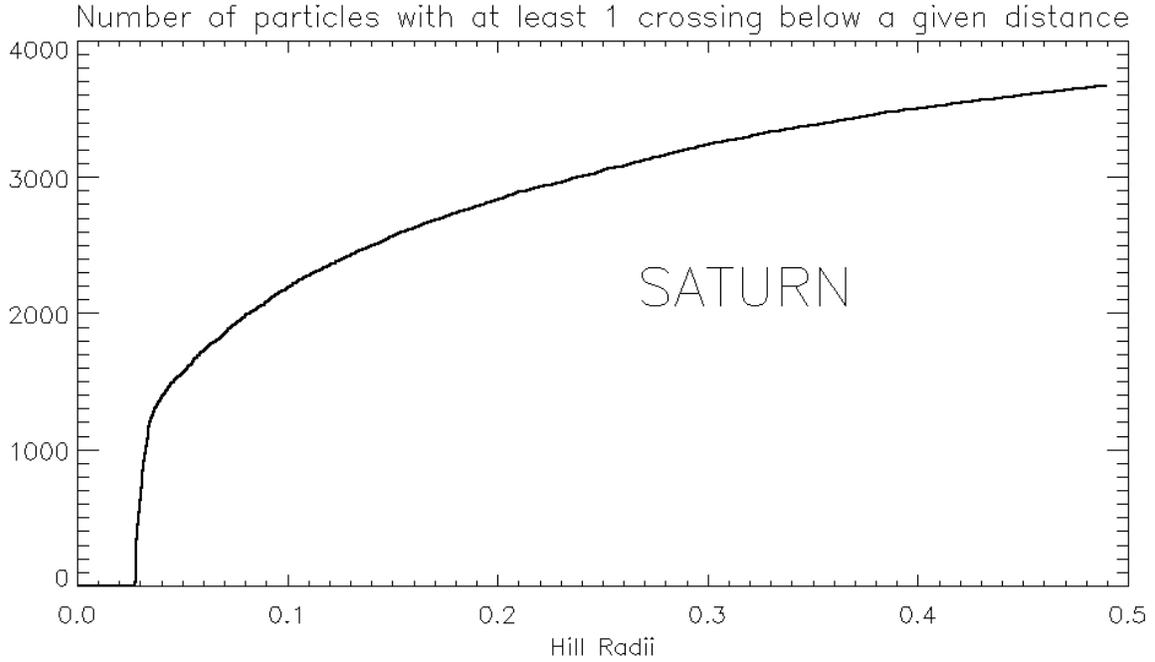}}
\caption{{\small Number of particles with at least one crossing inside a given
distance from Saturn. Simulations involved 5000 particles with total mass of $10$
M$_\oplus$. We estimate that roughly $\sim 5$ M$_\oplus$ pass within
0.2 R$_H$.}}
\label{fig:satcap}
\end{figure}

We follow the dynamical evolution of a $10$ M$_\oplus$ 
mass disk (where $M_\oplus$ is an Earth mass) of $5000$ 
particles spread between $4$ and $8$ AU
starting on low eccentricity and 
inclination orbits. Jupiter is placed at $5$ AU and Saturn is at $7.5$ AU (packed
configuration), both with low values of eccentricity and inclination. In Figures
\ref{fig:jupcap} and \ref{fig:satcap}
we plot the number of particles that crossed at least once within a given distance
of Jupiter and Saturn in $10^5$ years.
Not surprisingly, a large fraction
of the total number of particles in the simulation ($5000$) 
cross at least once within $0.1$ of the Hill sphere of
Jupiter or Saturn in the first $\sim 10^4$ yrs. 
This is because particles are scattered
during close encounters with the giant planets. 
Several caveats apply to these results. For one, collisional
fragments, which are easier to deliver to the 
circumplanetary disk, may collide
frequently, which would change their dynamics. 
This process is not modeled in
the results discussed above, and is left for future work. 
But we stress that collisions cannot prevent collisional debris from 
crossing the circumplanetary disk  of
either giant planet -- particles of any 
size at the gap edge and beyond are able to penetrate the 
Hill sphere of the giant planet during close
encounters. In some regimes of interest, 
such as cases in which nebula gas drag and 
collisional damping is included and 
meter-sized particles play an important role, 
one needs to treat the full fragmentation-coagulation 
problem. However, it is presently unclear how much mass
the tail end of the mass-spectrum contains. 

We calculate the total captured mass of the ice and rock components $M_j(X)$
delivered within a radius $X=R/R_P$ to the 
circumplanetary disk in time $\tau$ using (e.g., Estrada and Mosqueira 2006)

\begin{equation}
\begin{split}
M_j(X)= 2\pi\tau R_P^2 \int_{1}^{X} F_j(X^\prime)X^\prime\,dX^\prime;\,\,\,\,\,\,\,\,\,\,\,
\,\,\,\,\,\, \\
F_j(X) = v_0(1 + l\Theta/X)\int_{m_{min,j}}^{m_{max,j}} c_j\Phi_j(X,m)
m^{1-p}\,dm
\end{split}
\label{equ:massd},
\end{equation}

\noindent
where $F_j$ is mass inflow deposition rate per unit area of component $j$,
$v_0$ is the speed at infinity,
$p = (1/3)(q+2)$, $m_{min,j}$ and $m_{max,j}$ are the lower and upper bounds
of the mass distribution of component $j$, $\Theta =  GM_P/R_P v_0^2$ is the Safronov parameter,
$l \sim 1 - 2$ is a constant,
and $\Phi_j(X,m)$ is the fraction of a planetesimal with 
mass $m$ that is delivered to the disk at $X$, as obtained from
Eqs. \ref{equ:mdvdt} and \ref{equ:dmdt}. The $c_j$ are normalization coefficients
given by
 
\begin{equation}
c_{j} = \frac{(2-p)M_{pl}}{2\pi R_P^2v_0\tau}\left[\frac{1}{2}(X_D^2-1)+
l\Theta(X_D-1)\right]^{-1} \times
\begin{cases}
\frac{1-x_r}{m^{2-p}_{max,i} -m^{2-p}_{min,i}}  &\,{\rm{for}}\,\,{\rm{ice}};\\
\frac{x_r}{m^{2-p}_{max,r} -m^{2-p}_{min,r}}  &\,{\rm{for}}\,\,{\rm{rock}}\\
\end{cases},
\label{equ:ccoef}
\end{equation}

\noindent
where $M_{pl}$ is the mass of planetesimals passing through the disk, $x_r$ is the mass 
fraction of rock, and for our disk size we take $X_D = 0.2 R_H/R_P$. The surface density of components $j$,
$\Sigma_j$, solids-to-gas-ratio, and the ice mass fraction as a function of radius 
are obtained from the $F_j(X)$.

Pending more detailed
scattering/fragmentation simulations, we
assume a power-law mass distribution with 
slope $q = 3.5$ and an upper and lower particle size
cut-offs of $r_{max}= (3m_{max}/4\pi\rho_p)^{1/3} =200$ km and 
$r_{min}= (3m_{min}/4\pi\rho_p)^{1/3} = 0.001$ km for both rock and ice.
We allow for a differentiated
population of icy and rocky fragments with
a $30/70$ ratio by mass. Based on our $N$-body simulations, we estimate that about
$M_{pl} \sim 7$ M$_\oplus$ cross at least once within $0.2$ R$_H$. It should be
noted that particles $\sim 1$ km that cross
the disk at $\gtrsim 0.1$ R$_H$ are not in general ablated so they may
cross again within $0.2$ R$_H$ before being ejected. We take
the Safronov parameter $\Theta = 25$ representative
of a fairly cold population. The factor $l$ accounts for the different geometry of a flux of impactors through
the subnebula rather than the planet itself (Morfill et al. 1983; 
Cuzzi and Estrada 1998; Charnoz et al. 2009). The value for $l$ should be obtained from
numerical simulations. For now
we use $l = 2$, but the results are not sensitive to this parameter. 
The background nebula temperature is
set to $130$ K. For our subnebula parameters, we choose $\Sigma_a = 2\times 10^4$ g cm$^{-2}$,
$\Sigma_b = 5\times 10^3$ g cm$^{-2}$, $R_a = 15$ R$_J$, $R_b = 25$ R$_J$, and $n=1$.
The velocity of the planetesimal is given by $v = \sqrt{v_0^2 + v_{esc}^2}$.

\begin{figure}[t!]
 \resizebox{\linewidth}{!}{%
 \includegraphics{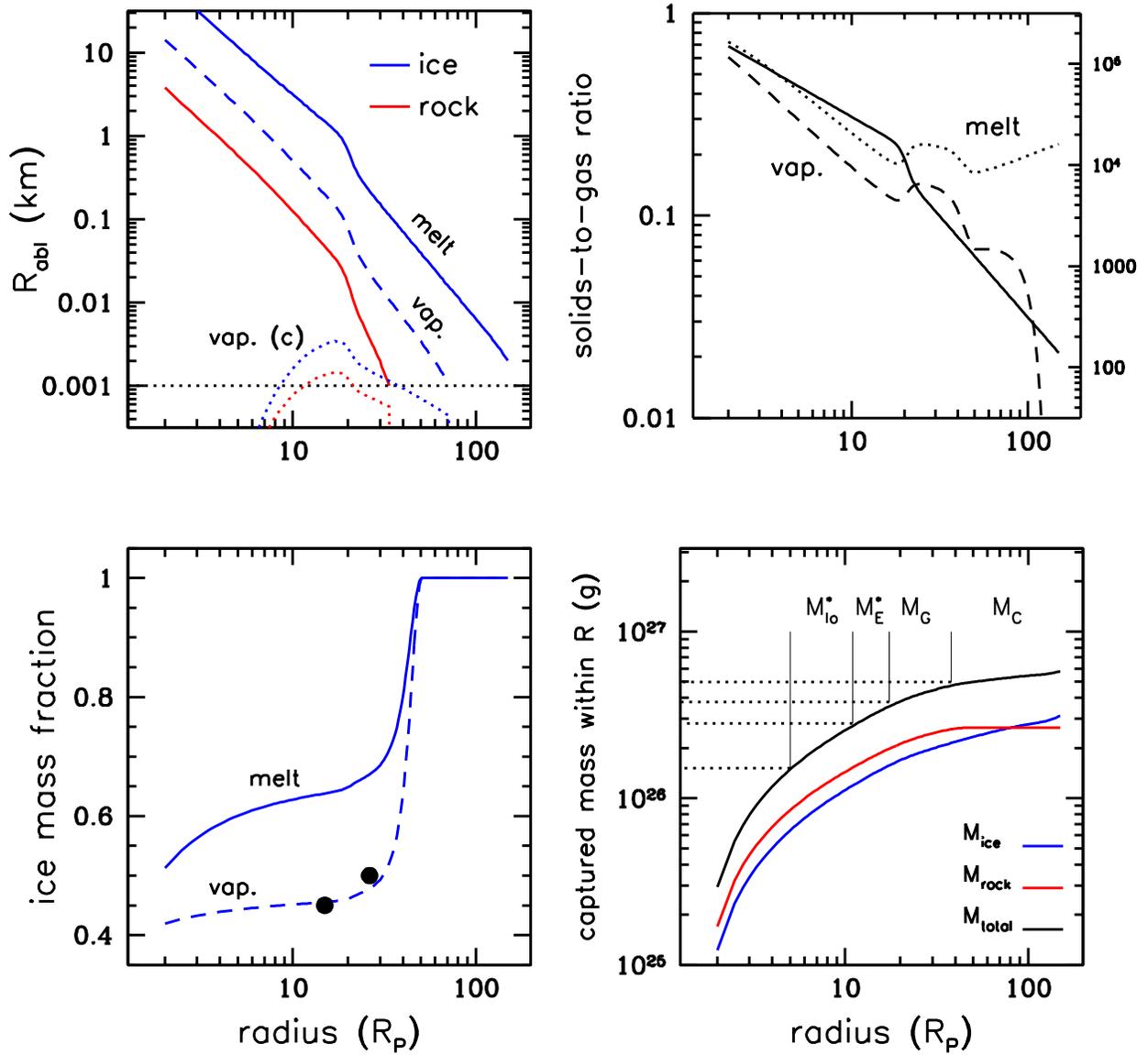}}
\caption{{\small Left upper panel: The planetesimal size
that gets ablated (solid and dashed curves) and captured
(dotted curves) as function of radial distance. The dotted horizontal
line corresponds to the lower size cut-off.
Right upper panel: The solids to gas ratio for the two ice models (dotted and dashed
curves) as a function of radial distance from Jupiter. The bold solid curve indicates the
gas surface density of the disk in g cm$^{-2}$ (right side). 
Left lower panel: The ice to rock ratio as a function
of radial distance. The solid dots
correspond to Ganymede (inner disk) and Callisto (outer disk). 
The ice mass ratio of solar nebula
planetesimal fragments crossing the disk is taken to be $0.3$. 
Lower right panel: Plot of the
mass delivered as a function of distance from Jupiter.
The two models correspond to melting and vaporizing ice.}}
\label{fig:jall}
\end{figure}

In Fig. \ref{fig:jall} we show the 
mass delivery by ablation and gas drag capture
of planetesimal fragments crossing the SEMM circumplanetary disk. 
The top left panel indicates the size of the largest particle 
that will be delivered to the disk in full by a combination of
ablation and capture of the leftover object (dotted curves).
It can be seen that
a higher proportion of mass is delivered by capture for rock than
for ice (especially for the ice melt curve). 
This is because less ablation makes the rock particle 
cross section and gas drag force larger for a given size.
The top right panel corresponds to the solids/gas fraction
that results from a cold population of planetesimal
fragments as a function of position in the disk. The surface gas
density is also indicated. Both the dotted and the
dashed curves exhibit a bump during the inner to outer disk
transition. This jump in the concentration of solids occurs because in the transition region
the surface gas density of the disk decreases faster than the decrease in the mass
delivery. Consequently, the concentration of solids
increases. The subsequent outer disk rise in the dotted curve and bump
in the dashed curve are related to the $1/R^2$ surface density profile in the
outer disk (compared to a $1/R$ profile in the inner disk) together with the
transition to an isothermal temperature profile in the outer disk (compared
to a $1/R$ profile in the inner disk). The dashed curve drops towards
the outer edge because the temperature of the planetesimal becomes too low
for significant vaporization either of rock or
ice (whereas the temperature is still high
enough to melt water-ice). The bump in the concentration of solids at the
transition between the inner and outer disk is tantalizing in that
it can help to accrete satellites close to the present location
of Ganymede (and Titan for Saturn's disk). But it is premature to say that
this feature is significant. The bottom left panel
indicates the resulting ice/rock fraction of the disk. 
The bottom right panel indicates the total
mass delivered within a radial distance $R$. The region between
the bars contain enough mass to make each of the indicated satellites (reconstituted in the
case of Io and Europa). The mass delivered was obtained by using
the ice melt results for $T_{surf} < T_c$ and the vaporization results for
larger $T_{surf}$. These results indicate that enough mass can be delivered
to account for the reconstituted Galilean satellites. But we stress that
the mass delivered to the disk is dependent on the parameters chosen and the
capture condition $v_{final} = 0$. On the other
hand, the capture condition is a conservative choice 
in that significantly more mass will be delivered
to the circumplanetary disk than we assume here. We reserve
a parameter study for later work.

The two ice curves shown correspond to ablation by
melting and vaporization in Eq. \ref{equ:dmdt}. As one would expect,
more ice is delivered when ablation turns on
at the lower temperature of melting $T_m$ versus
vaporization $T_c$. The good agreement with the
mass budget and compositions of 
Ganymede and Callisto would appear
to argue in favor of the vaporization curves. However, this
conclusion is premature for several reasons. First, these 
calculations make the assumption that the
population of disk crossers is completely fractionated, which is
unrealistic even if primordial planetesimals were
fully differentiated (see Sec. 4), as some mixing would be expected during
collisional disruption. A more realistic treatment 
would lower the ice/rock ratio of the entire disk. Two, vaporization
of water-ice actually starts below the melting temperature
$T_m = 273 ^o$ K (whereas for rock temperatures $200 ^o$ K above
melting $T_m = 1800 ^o$ K are needed 
before significant vaporization takes place; 
Podolak et al. 1988). Third, the total mass delivered to the circumplanetary
disk is dependent on parameter choices.
However, the result that the outer disk is icier
than the inner disk is independent of the specifics of the 
ablation model. In fact, given a fully fractionated population crossers, 
the outer disk receives material mostly by ablation of
ice, and essentially no rock either by ablation or capture. 
Also, these results lead one to expect 
that the circumplanetary disk is enriched in ice over solar mixtures
for a broad range of parameter choices.

\begin{figure}[t!]
 \resizebox{\linewidth}{!}{%
 \includegraphics{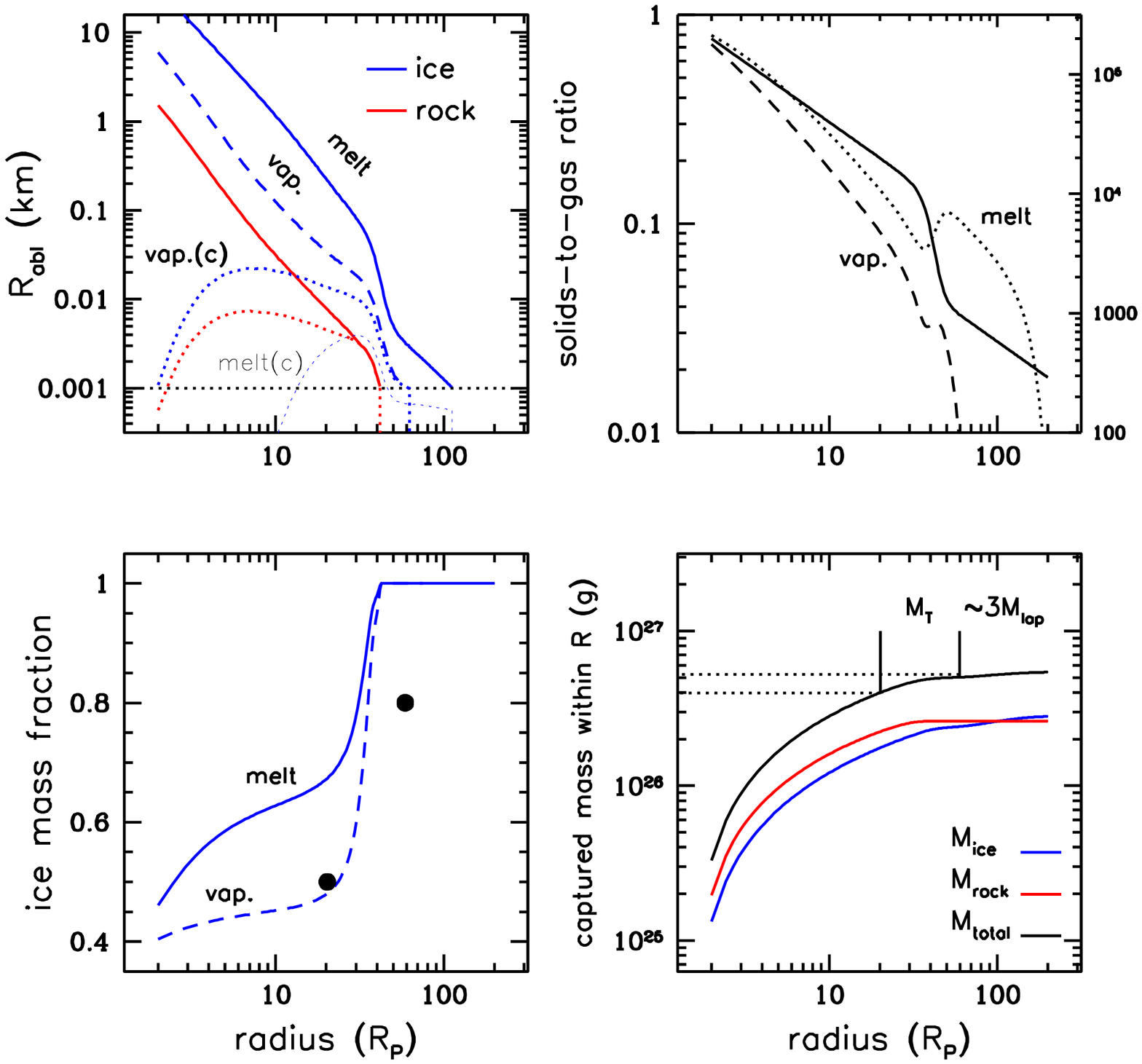}}
\caption{{\small Left upper panel: The planetesimal size
that gets ablated (solid and dashed curves) and captured
(dotted curves) as function of radial distance. The dotted horizontal
line corresponds to the lower size cut-off.
Right upper panel: The solids to gas ratio for the two ice models (dotted and dashed
curves) as a function of radial distance from Saturn. The bold solid curve indicates the
gas surface density of the disk in g cm$^{-2}$ (right side). 
Left lower panel: The ice to rock ratio as a function
of radial distance. The solid dots
correspond to Titan (inner disk) and Iapetus (outer disk). 
The ice mass ratio of solar nebula
planetesimal fragments crossing the disk is taken to be $0.3$. 
Lower right panel: Plot of the
mass delivered as a function of distance from Saturn.
The two models correspond to melting and vaporizing ice. }}
\label{fig:sall}
\end{figure}

Similarly for Saturn, we
assume a power-law mass distribution with 
slope $q = 3.5$ and allow for a differentiated
population of icy and rocky fragments with
a $30/70$ ratio by mass. Based on our simulations,
we estimate that about
$M_{pl} \sim 5$ M$_\oplus$ cross at least once within $0.2$ R$_H$ of Saturn. 
We let $\Theta = 100$ and
an upper and lower particle size-cutoffs of $r_{max}=50$ km and
$r_{min}=0.001$ km for both rock and ice based on the larger fraction of rubble
and lower velocity dispersion expected for Saturn.
The background nebula temperature is
set to $90$ K. For our subnebula parameters, we chose $\Sigma_a = 2\times 10^4$ g cm$^{-2}$,
$\Sigma_b = 2\times 10^3$ g cm$^{-2}$, $R_a = 20$ R$_S$, $R_b = 58$ R$_S$, and $n=3$. As we can 
see from Fig. \ref{fig:sall}, rocky fragments in the $0.001 - 1$ km are not
either captured or ablated, whereas icy fragments are delivered mostly
by ablation. Thus, delivery of solids
via ablation of a non-homogeneous population of planetesimal fragments leads to
an enrichment of water-ice in Saturn's outer disk.
 
In summary, these results show that it is possible to deliver enough
mass to account for the regular satellites of 
Jupiter and Saturn.
Furthermore, ablation and the capture cross-section 
of meter-sized objects can
lead to ice/rock fractionation and account 
for the composition of Iapetus, as well as those of
Ganymede, Callisto and Titan. 
This is the only explanation currently available
for Iapetus' icy composition, 
and may also indicate that regular satellites 
overall may be depleted in rock
with respect to solar composition mixtures, 
which might explain why their densities are lower than that
of similar size Kuiper belt objects. 
We stress that while the amount of mass delivered to the
circumplanetary disks of Jupiter and Saturn is dependent
on parameter choices, such as the upper size cut-off and
the slope of the mass spectrum, the velocity dispersion
at infinity and the capture condition, we found that
the efficiency of delivery is high enough that the mass
budget of the satellites can easily be accommodated.
We find that the water-enrichment
of the outer disks of Jupiter and Saturn 
is a robust feature given a differentiated
population of icy and rocky interloper fragments. 

While full ice/rock fractionation is unlikely, partial
separation can account for Iapetus' icy 
composition. Such a population
results from the fragmentation of a partially differentiated, 
$^{26}$Al-heated first generation of
$\sim 10$ km planetesimals (Mosqueira et al. 2009). 
Indeed, it is likely that most of the
mass in the first generation of planetesimals resided in
objects $10-100$ km in the first $10^5-10^6$ years, so that
these objects incorporated significant
amounts of $^{26}$Al, leading to partial differentiation,
as we show below.

\begin{table}[ht]
\caption{Physical Parameters}
\centering
\begin{tabular} {l c c}
\hline
Parameter & Ice & Rock \\ [0.5ex]
\hline
$\mu$ (g mole$^{-1}$) & 18 & 50 \\
$\rho_s$ (g cm$^{-3}$) & 0.92 & 3.4 \\
$E_{0c}$ (erg g$^{-1}$) & 2.8 $\times$ 10$^{10}$ & 8.08 $\times$ 10$^{10}$ \\ 
$E_{0m}$ (erg g$^{-1}$) & 5.94 $\times$ 10$^9$ & 3.04 $\times$ 10$^{10}$ \\
$H_f$ (erg g$^{-1}$) & 3.33 $\times$ 10$^{9}$ & 1.43 $\times$ 10$^{10}$ \\
$T_m$ ($^o$K) & 273 & 1800 \\
$T_c$ ($^o$K) & 648 & 4000 \\ [1ex] 
\hline
\end{tabular}
\end{table}

\section{Differentiation of primordial planetesimals}

The thermal evolution of comets (Prialnik et al. 2008) and asteroids 
(Ghosh et al. 2006) powered by short-lived
radionuclides has been treated in an extensive number of publications,
and it is not possible for us to do it justice in this brief section. Our aim
here is to get a handle on the degree of differentiation of
$\sim 10$ km primordial planetesimals forming in the Jupiter-Saturn
region. This population of objects would have formed in a faster
timescale and in a hotter region of the nebula than their Kuiper belt 
counterparts, yet still outside the snowline. With the 
possible exception of $\sim 100$ km Phoebe, which may itself be at least partially
differentiated (Johnson et al. 2009), and whose composition appears
to match that of other outer solar nebula objects (Johnson and Lunine 2005),
there are no known examples of this population.

Our simplified treatment follows along the same lines as that
of other works only applied to our region of interest. We leave a more accurate
study for later work (in collaboration with M. Podolak). We solve the
energy balance equation for a body of size $r$ up to the melting temperature 
$T_m$ in a time $t_m$ (e.g., Jewitt et al. 2007)

\begin{equation}
c_p(T) \frac{dT}{dt} = x_r  \sum_{j} \tau_j \gamma_{0,j} H_j -
\frac{3 k(T) T}{r^2 \rho_p},
\label{equ:tmelt}
\end{equation}

\noindent 
where $\rho_p$ is the mean density, $x_r$ is the mass fraction of rock,
$\tau_j$, $\gamma_{0,j}$, and $H_j$ are the half-life, initial
abundance and decay energy per unit mass of radioisotope $j$,
and $T$ is the mean temperature. The conductivity $k(T)$ and specific heat  
$c_p(T)$ are given in terms of their respective values for rock ($r$) and ice ($i$)
by (e.g., Multhaup and Spohn 2007)

\begin{equation}
\begin{split}
k(T) = k_rf_r + (1-f_r)k_i(T) \\
c_p(T) = c_rx_r + (1-x_r)c_i(T)
\end{split},
\end{equation}

\noindent
where $f_r$ is the rock volume fraction. In addition, the conductivity and density
are adjusted by factors of $1 - \phi^{2/3}$ and $1-\phi$, respectively, to account
for porosity $\phi$ (Smoluchowski 1981). Equation \ref{equ:tmelt}
yields the temperature of the planetesimal subject to
radionuclide heating and conductive cooling. 
%
%
The relevant parameters for these calculations are given in Table 2.
 
\begin{figure}[t!]
 \resizebox{\linewidth}{!}{%
 \includegraphics{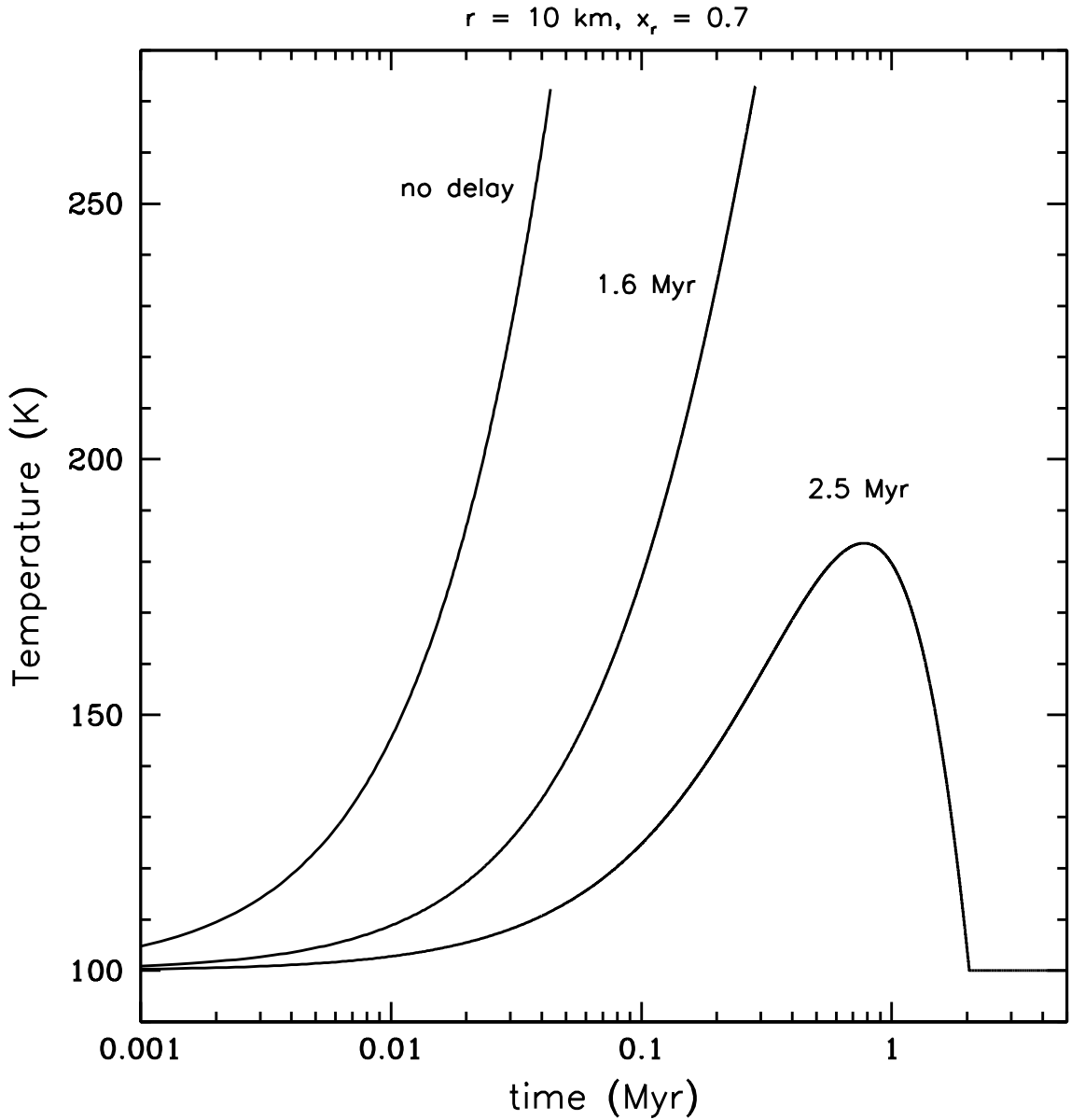}}%
\caption{{\small Mean planetesimal temperature as a function of
time. The curves correspond to different delay times between CAIs and the
formation time of the primordial $10$ km planetesimal. 
The rock fraction is $x_r = 0.7$ by mass. The porosity is $\phi = 0.3$. Since
planetesimals in the Jupiter-Saturn region 
form in the first $\sim 1$ Myr they 
are expected to reach melting temperatures.}} 
\label{fig:tmelt}
\end{figure}

A variety of outcomes are possible 
depending on poorly constrained model 
parameters, but melting is likely to take place if the object incorporates
a significant amount of $^{26}$Al. A study by
Merk and Prialnik (2006) that includes radioactive heating, accretion, 
transformation of amorphous to crystalline ice and melting of water 
ice in the formation of trans-neptunian objects with sizes $\sim 10$ km 
finds that the occurrence of liquid water may be common.
Since Kuiper belt objects form in longer timescale and accrete
in a colder environment, it is not surprising that planetesimals 
forming early-on in the Jupiter-Saturn region are prone to melting.
In Fig. \ref{fig:tmelt} we show the mean temperature as function of
time for several values of the delay between CAIs and planetesimal
formation (assuming $^{26}$Al is homogeneously distributed
out to Saturn). The nebula temperature is $100$ K.
The values for the physical parameters used in this
model are given in Table 1.
We start with crystallized ice (and no CO), due to our higher temperature
of accretion. We ignore serpentinization reactions, 
which would lead to further heating.
Melting of these objects is difficult
to avoid in the Jupiter-Saturn region even with significant
delays in the planetesimal formation time.
The reason is that the energy supply of $^{26}$Al
greatly exceeds the amount of energy required for melting the entire
object. For a planetesimal of size $\sim 10$ km, conduction
cannot remove this energy in time,
leading to at least partial differentiation. 
This also fits with more detailed studies of
hydrological activity of small icy planetesimals in the asteroid belt (Young 2001), where
the background temperature is higher.
The disruption of partially differentiated planetesimals would then lead to a 
population of icy/rocky fragments available to ablate through 
the extended Kronian gas disk.
 
Above we neglect heat transport by water vapor flow through a porous medium 
driven by internal sublimation (Prialnik and Podolak 1995).
Yet, the presence of liquid water is not required for partial 
differentiation to take place. Enhanced conductivity 
by water vapor flow through porous media may discourage melting, 
but would still result in a layered internal state as the core would 
become water-depleted (Prialnik et al. 2008). To check this we
we add a cooling term due to the rate of vapor sublimation given by
(e.g., Prialnik and Podolak 1995)

\begin{equation}
Q_v = S_\phi H_v ({\cal P(T)} - P_v) \left( \frac{\mu_{ice}}{2 \pi R_g T} \right)^{1/2}
\end{equation}

\noindent
where $S_\phi = - (3/r_0) \phi$ ln $\phi$ is the surface to volume ratio
(Koponen et al. 1997), $r_0$ is the grain size,  
$H_v$ is the latent heat of vaporization,
${\cal P(T)}$ is the saturated vapor pressure, $P_v$ is the
gas pressure, $\mu_{ice}$ is the molecular
mass of ice, and $R_g$ is the gas constant. The value for the
porosity is uncertain and the sublimation rate is model-dependent. 
We treat it as an adjustable parameter and find 
that for values relevant to the Jupiter-Saturn region
with nebula temperature $\sim 100$ K and significant
quantities of $^{26}$Al one must sublimate
a significant fraction of the water-ice of the object in order
to keep it from melting. Thus, whether 
sublimation is sufficiently effective to preclude
melting or not, radiogenic heating is likely to result in
separation of rock and water in our regime
of interest where nebula temperatures are higher and
accretion timescales shorter than in the Kuiper belt, which
leads us to expect that the primordial population
of planetesimals in the Jupiter-Saturn region is partially
differentiated.

\begin{table}[ht]
\caption{Thermal evolution parameters}
\centering
\begin{tabular} {l}
\hline
Thermal conductivity of ice$^a$\\ [0.5ex]
$k_i = 4.8812 \times 10^{7}/\,T + 0.4685 \times 10^5$ (g cm s$^{-3}$) \\ [0.5ex]
Thermal conductivity of water\\ [0.5ex]
$k_w = 0.55 \times 10^5$ (g cm s$^{-3}$) \\ [0.5ex]
Thermal conductivity of rock$^b$\\ [0.5ex]
$k_r = 4.2 \times 10^5$ (g cm s$^{-3})$ \\ [0.5ex]
Specific heat of ice$^a$ \\ [0.5ex]
$c_i = 7.037 \times 10^4 \,\,T + 1.85 \times 10^6$ (cm$^2$ K$^{-1}$ s$^{-2}$) \\ [0.5ex]
Specific heat of water \\ [0.5ex]
$c_w = 4.19 \times 10^7$ (cm$^2$ K$^{-1}$ s$^{-2}$) \\ [0.5ex]
Specific heat of rock$^b$ \\ [0.5ex]
$c_r = 1.2 \times 10^7$ (cm$^2$ K$^{-1}$ s$^{-2}$) \\ [0.5ex]
Half life of $^{26}{\rm{Al}} = 7.2 \times 10^5$ yrs\\ [0.5ex]
Initial abundance of $^{26}{\rm{Al}} = 7.0 \times 10^{-7}$ from
$\frac{^{26}{\rm{Al}}}{^{27}{\rm{Al}}} = 5 \times 10^{-5}$ \\ [0.5ex]
Heat production rate$^c$ of $^{26}{\rm{Al}} = 1.16 \times 10^{17}$ (erg g$^{-1}$) \\ [1.0ex]
\hline
$^a$Hobbs (1974); $^b$Ellsworth and Schubert (1983); $^c$Castillo-Rogez et al. (2009)
\end{tabular}
\end{table}

\section{Conclusions}

There are two self-consistent models of satellite formation
(MEa,b; Estrada and Mosqueira 2006; see Estrada \etal 2009 for a review).  
Both of these models
treat planetesimal dynamics explicitly and derive the solids
leading to the formation of the regular satellites from the ``shower''
of planetesimal fragments following the 
formation of the giant planets. However, only one
of the two accretes satellites in the presence of a relatively massive (SEMM)
gas component (MEa,b; see Mosqueira \etal 2009 for a review) able
to ablate and capture a significant fraction of the planetesimal rubble
crossing the circumplanetary disk. 
Simply put, there are two basic options available to modelers: 
first, guided by the distribution of satellite mass
for Jupiter and Saturn (especially but not exclusively the large 
separation between Titan and Iapetus),
one can infer a {\it gaseous} subnebula with a 
relatively dense inner region and a
long tail out to the location of the irregular satellites; second, assuming
instead that the gas component is dissipated by an {\it unknown}
mechanism in a short timescale, one can consider a {\it gas-poor} 
satellite formation model. 

Here we show that it is possible
to understand the mass budgets and compositions of the regular
satellites of the giant planets in terms of a two-component
gaseous model\footnote{A two component model has the added benefit of
providing a natural explanation for the Callisto-Ganymede
dichotomy without resorting to fine-tuning model
parameters (see Estrada et al. 2009 and Mosqueira et al. 2009).
This also fits with the observation that the composition of
Callisto is icier than that of Ganymede.}.
In particular, the result that (even allowing for porosity)
Iapetus is water-ice rich compared to
solar mixtures provides strong support for the presence of a
two-component (SEMM) gas disk at the time of satellite formation.
Indeed, a lower surface density outer disk is {\it required}
to simultaneously satisfy the constraints imposed by
Iapetus' small mass and ice-rich composition (compared
to Callisto, Ganymede and Titan). Thus, Iapetus' 
distance from the primary sets it apart from
all other regular satellites.  But we stress that our framework applies
in general to the formation of regular satellites around gaseous
giants. In fact, our results also account for the water-ice
enrichment of Callisto compared to Ganymede as observed.

For our model to apply, the first generation of planetesimals
in the Jupiter-Saturn region must be at least partially differentiated.
We conduct a preliminary investigation of the
thermal evolution of planetesimals, and find that $10$ km bodies
in Jupiter-Saturn region are more likely to melt and differentiate than
similar composition and size objects located in the
Kuiper belt region because of the shorter accretion times
and hotter nebula environment closer to the Sun.
Since we argue that the first generation
of planetesimals in the Jupiter-Saturn region 
form quickly and incorporate
significant quantities of $^{26}$Al, these
objects are at least partially differentiated. As a result, the
collisional cascade following giant planet formation
leads to the production of a non-homogeneous
population of planetesimal fragments
available to ablate through the circumplanetary disks 
of the giant planets, resulting in water-ice
enrichment in the outer disk. Given the small
degree of radial mixing expected in the quiescent-disk
SEMM satellite formation model (MEa,b), a radial
compositional gradient can be preserved in the bulk
properties of the regular satellites as observed.

We do not include the abundances of other volatile species, 
as we do not have good constraints for
their contribution to the bulk compositions of either Kuiper
belt objects or regular satellites. Also, we do not
consider a phase of condensed, refractory carbon (in contrast
to CH$_4$, which is trapped in water-ice).
Solid carbon affects the condensate density and reduces the
amount of CO available to sequester oxygen 
(Johnson and Lunine 2005; Wong et al. 2008).
However, there is presently no evidence that such a phase plays
a significant role in the bulk compositions of the regular satellites, 
and no reason for including it in Iapetus but not doing so in other 
regular satellites, Phoebe, or Kuiper belt objects.
Although a number of the parameters employed here are at 
present uncertain, our work ties regular
satellite formation to the outer solar nebula, 
so that constraints from observational studies 
of the compositions of the giant
planets, or comet populations, and further
theoretical work of the requirements for giant planet core
formation, or Kuiper belt constraints, can be used 
in connection with the regular satellites as well. In
fact, this is the first study to establish a direct link between
solar nebula planetesimals and the bulk properties of
regular satellites, thus
opening up an entirely new research direction. A
number of effects have been left for further work, notably: 
collisional damping in the planetesimal dynamics
following giant planet formation, and the effect of multiple
disk crossings for those objects that are not captured during
the first pass. Also, our results assume that the 
size distribution and size cut-offs,
and velocity at infinity of rocky and icy fragments are the 
same.

However, we have tested parameter sensitivities and
find that the result that the outer disk
of Saturn is ice-rich compared to its inner disk, and
also compared to the outer solar nebula, is robust, provided that
primordial planetesimals are at least partially differentiated. 
The reason is that delivery of solids in the outer disk is dominated 
by melting and vaporization
of water-ice at temperatures of hundreds of degrees Kelvin.
This result is a consequence of the lower gas surface density
and lower planetesimal velocities in the outer disk. Consequently,
icy fragments of size $\gtrsim 1$ m are delivered by ablation 
(and some capture), whereas delivery of rocky fragments of similar
size is highly inefficient. This is mainly 
because the temperature of objects crossing the 
outer disk is not high enough for
rock to melt or vaporize. Additionally, overall a higher proportion
of the rock content is delivered by capture, which
drops off as the surface density gets lower. 
This is because for less pronounced
ablation the cross section of the traversing planetesimal and
gas drag force are larger, resulting in a higher proportion
of mass captured compared to ablated.
Also, rocky objects are more difficult to capture for a given object
size due to their greater density but this may
be compensated by the availability of somewhat
smaller rocky fragments.
 
Still, much work remains to be done to pin down the
degree of enrichment quantitatively; in particular,
while second pass-capture of rocky fragments does not
erase a subnebula compositional gradient signature, it can
deliver rock to the disk via capture
and should be explored. Also, while collisional
damping of planetesimal eccentricities does not
prevent delivery of collisional rubble to the circumplanetary
disk during close encounters with the giant planet,
the amount of mass delivered is likely to be affected,
as it depends on the mass spectrum. Moreover, other
delivery mechanisms, such as collisional
capture and three-body effects, may also contribute to the
mass budget of the regular satellites but have not
been included in our analysis. Finally, we
assume a starting condition of a gas disk without
solid content. Yet planetesimals dissolve in the gas giant 
envelope, thus enhancing the planet's metallicity (Pollack et al. 1986)
and possibly the gas disk as well. 
Most of this mass delivery 
takes place before the mass of the gaseous 
envelope grows larger than that of the core. 
This is then followed by a {\it dilution} during 
runaway gas accretion (Pollack et al. 1996).
It is unclear whether this source of solids will contribute
significantly to the regular satellites, especially to those
located far from the planet. We expect that
the outer disk forms {\it after} envelope collapse,
as the giant planets open a gap in the nebula, and that
this gas is relatively free of solids (see Estrada et
al. 2009 for further discussion). Moreover, dust settling in 
the envelope of the giant planet may also limit the solid content that
is left behind in the subnebula far from the planet. Ice/rock fractionation
is also possible. We leave this issue for further work.

{\it Acknowledgements}. We would like to thank Jay Melosh
and Uma Gorti for useful discussions. I. Mosqueira's research
is supported by NASA PG\&G, and OPR grants. P. Estrada contribution is
supported by a NASA OSS grant.

\newpage
\begin{center}
{\bf References}
\end{center}
\begin{list}{}{\leftmargin \parindent \itemindent -\parindent \itemsep 0in}



\item Anderson, J. D., Johnson, T. V., Schubert, G., Asmar, S,
Jacobson, R. A., Johnston, D., Lau, E. L., Lewis, G., Moore, W. B.,
Taylor, A., Thomas, P. C., and G. Weinwurm 2005. Amalthea's density is less
than that of water. {\it Science}, {\bf 308}, 1291-1293.


\item Atreya, S. K., Wong, M. H., Owen, T. C.,
Mahaffy, P. R., Niemann, H. B., de Pater, I., Drossart, P., Encrenaz, T.,
1999. A comparison of the atmospheres of Jupiter and Saturn:
deep atmospheric composition, cloud structure, vertical mixing, and origin.
{\it Plan. and Space Sci.} {\bf 47}, 1243-1262.

\item Ayliffe B., and Bate M. R., 2009. Circumplanetary disc properties obtained 
from radiation hydrodynamical simulations of gas accretion by protoplanets.
{\it M.N.R.A.S} {\bf 397}, 657-665.
	
\item Bai, X., Goodman, J., 2009.
Heat and Dust in Active Layers of Protostellar Disks.
{\it Astrophys. J.} {\bf 701}, 737-755.

\item Baker, J., Bizzarro, M., Wittig, N., Connelly, J., Haack, H., 2005.
Early planetesimal melting from an age of 4.5662Gyr for differentiated meteorites.
{\it Nature} {\bf 436}, 1127-1131.


\item Bate, M. R., Lubow, S. H., Ogilvie, G. I., and K. A. Miller 2003.
Three-dimensional calculations of high- and low-mass planets embedded in
protoplanetary discs. {\it M.N.R.A.S.}, {\bf 341}, 213-229.


\item Beckwith, S. V. W., Henning, T., and Nakagawa, Y., 2000. Dust
properties and assembly of large particles in protoplanetary disks.
 In {\it Protostars and Planets IV}. Eds., V. Mannings, A. P. Boss, and 
 S. S. Russell, University of Arizona Press, 533-558.

\item Bell, K. R., Cassen, P. M., Wasson, J. T., Woolum, D. S., 2000. The
FU Orionis phenomenon and solar nebula material. In {\it Protostars and
Planets IV}. Eds., V. Mannings, A. P. Boss, and S. S. Russell, University of
Arizona Press, 897-926.


\item Benz, W., and E. Asphaug 1999. Catastrophic disruptions revisited.
{\it Icarus}, {\bf 142}, 5-20.

\item Birnstiel, T., Dullemond, C. P., Brauer, F., 2009.
Dust retention in protoplanetary disks. {\it Astron. \& Astrophys.}
In press.

\item Bodenheimer, P., and J. B. Pollack 1986. Calculations of the
accretion and evolution of the giant planets: The effects of solid cores.
{\it Icarus} {\bf 67}, 391-408.


\item Bottke, W. F., Durda, D. D., Nesvorny, D., Jedicke, R., Morbidelli, A., 
Vokrouhlicky, D., Levison, H. F., 2005. Linking the collisional history of the 
main asteroid belt to its dynamical excitation and depletion.	
{\it Icarus} {\bf 179}, 63-94.

\item Brauer, F., Henning, Th., Dullemond, C. P., 2008.
Planetesimal formation near the snow line in MRI-driven turbulent protoplanetary disks.
{\it Astron. \& Astrophys.} {\bf 487}, L1-L4

\item Bronshten, V. A., 1983. Ablation of meteoroids. {\it Meteo. Iss.}
{\bf 8}, 38-50. In Russian.


\item Brown, M. E., Schaller, E. L., 2007.
The Mass of Dwarf Planet Eris. {\it Science} {\bf 316},
1585.



\item Buie, M. W., Grundy, W. M., Young, E. F., Young, L. A., Stern, S. A., 2006.
Orbits and Photometry of Pluto's Satellites: Charon, S/2005 P1, and S/2005 P2.
{\it Astron. J.} {\bf 132}, 290-298.

\item Buratti, B. J., Hicks, M. D., Lawrence, K. J., 2009. 
Movement of Low-albedo Dust from Small Irregular Satellites 
to the Main Satellites of the Outer Planets.
{\it BAAS} {\bf 41}.


\item Canup, R. M., W. R. Ward 2002. Formation of the Galilean Satellites: 
Conditions of Accretion. {\it Astron. J.} {\bf 124}, 3404-3423.

\item Cassen, P., and Pettibone, D. 1976. Steady accretion of a 
rotating fluid. {\it Astrophys. J.} {\bf 208}, 500-511.

\item Castillo-Rogez, J. C., Matson, D. L., Sotin, C., Johnson, T. V., 
Lunine, J. I., Thomas, P. C., 2007. Iapetus' geophysics: Rotation rate, 
shape, and equatorial ridge. {\it Icarus} {\bf 190}, 179-202.

\item Castillo-Rogez, J., Johnson, T. V., Lee, M. H., Turner, N. J., Matson, 
D. L., Lunine, J., 2009. $^{26}$Al decay: Heat production and a revised age for Iapetus.
{\it Icarus}. In press.

\item Chabot, N. L., Haack, H., 2006.
Evolution of Asteroidal Cores.
In {\it Meteorites and the Early Solar System II}. Eds., D. S. Lauretta, and 
H. Y. McSween Jr., University of Arizona Press, 747-771.

 
\item Charnoz, S., and A. Morbidelli 2003. Coupling dynamical and collisional 
evolution of small bodies: an application to the early ejection of 
planetesimals from the Jupiter-Saturn region. {\it Icarus}, {\bf 166}, 141-156.

\item Charnoz, S., Morbidelli, A., 2007. Coupling dynamical and collisional evolution 
of small bodies. II. Forming the Kuiper belt, the Scattered Disk and the Oort Cloud.
{\it Icarus}{\bf 188}, 468-480.

\item Charnoz S., Morbidelli, A., Dones, L., and Salmon, J. 2009. Did Saturn's rings 
form during the Late Heavy Bombardment? {\it Icarus} {\bf 199}, 413-428.








\item Currie, T., Lada, C. J., Plavchan, P., Robitaille, T. P., Irwin, J., Kenyon, S. J., 2009.
The Last Gasp of Gas Giant Planet Formation: A Spitzer Study of the 5 Myr Old Cluster NGC 2362.
{\it Astrophys. J.} {\bf 698}, 1-27.


\item Cuzzi, J.N.; Estrada, P.R., 1998. Compositional evolution of Saturn's rings due to meteoroid 
bombardment. {\it Icarus} {\bf 132}, 1-35. 

\item Cuzzi, J. N., Hogan, R. C., Shariff, K., 2008.
Toward Planetesimals: Dense Chondrule Clumps in the Protoplanetary Nebula.
{\it Astrophys. J.} {\bf 687},	 1432-1447.

\item Cuzzi, J. N., and Weidenschilling S., 2006. Particle gas dynamics and
primary accretion. In {\it Meteors and the Early Solar System II}. Eds.,
D. Lauretta, L. A. Leshin, and H. McSween, University of 
Arizona Press, 353-381.
	
\item Chyba, C. F., 1993.
Explosions of small Spacewatch objects in the Earth's atmosphere.
{\it Nature} {\bf 363}, 701-703.




\item Davis, D. R., Farinella, P., 1997.
Collisional Evolution of Edgeworth-Kuiper Belt Objects.
{\it Icarus} {\bf 125}, 50-60.

\item Dodson-Robinson, S. E., Bodenheimer, P., Laughlin, G., Willacy, K., 
Turner, N. J., Beichman, C. A., 2008. Saturn Forms by Core Accretion in 3.4 Myr.
{\it Astrophys. J.} {\bf 688}, L99-L102

\item Dohnanyi, J. W., 1969. Collisional model of asteroids and their
debris. {\it J. Geophys. Res.} {\bf 74}, 2531-2554.
	
\item Dominik, C., Blum, J., Cuzzi, J. N., Wurm, G., 2007.
Growth of Dust as the Initial Step Toward Planet Formation.
In {\it Protostars and Planets V}. Eds., B. Reipurth, D. Jewitt, and K. Keil, 
University of Arizona Press, Tucson, 783-800.

\item Dominik, C., Dullemond, C. P., 2008.
Coagulation of small grains in disks: the influence of residual infall and 
initial small-grain content. {\it Astron. \& Astrophys.} {\bf 491}, 663-670.


	
\item Dullemond, C. P., Dominik, C., 2005. 
Dust coagulation in protoplanetary disks: A rapid depletion of small grains
{\it Astron. \& Astrophys.} {\bf 434}, 917-986.

\item Durham, W. B., McKinnon, W. B.; Stern, L. A., 2005.
Cold compaction of water ice. {\it Geophys. Res. Let.} {\bf 32}	
L18202.

\item Ellsworth, K., and Schubert, G., 1983. Saturn's icy satellites: thermal and
structural models. {\it Icarus} {\bf 54}, 490-510.




\item Estrada, P. R., and I. Mosqueira 2006. A gas-poor planetesimal
capture model for the formation of giant planet satellite systems.
{\it Icarus} {\bf 181}, 486-509.

\item Estrada, P. R., I. Mosqueira, J. J. Lissauer, G. D'Angelo, D. P. Cruikshank 2009.
Formation of Jupiter and Conditions for Accretion of the Galilean Satellites.
To appear in the Europa Book, Eds. W. McKinnon, R. Pappalardo, and
K. Khurana. University of Arizona Press.


	
\item Ghosh, A., Weidenschilling, S. J., McSween, H. Y., Jr., Rubin, A., 2006.
Asteroidal Heating and Thermal Stratification of the Asteroidal Belt.
In {\it Meteorites and the Early Solar System II}. Eds., D. S. Lauretta and 
H. Y. McSween Jr., University of Arizona Press, 555-566.











\item Hayashi, C., 1981. Structure of the solar nebula, growth and
decay of magnetic fields, and effects of magnetic and turbulent
viscosities on the nebula. {\it Prog. Theor. Phys. Suppl.},
{\bf 70}, 35-53.

\item Hobbs, P. V., 1974 {\it Ice Physics}, Oxford University Press, London. pp.
347-363.



\item Housen, K. R., Schmidt, R. M., Holsapple, K. A., 1991. Laboratory
simulations of large-scale fragmentation events. {\it Icarus}
{\bf 94}, 180-190.


\item Hubickyj, O., Bodenheimer, P., and J. J. Lissauer 2005. Accretion
of the gaseous envelope of Jupiter around a $5-10$ Earth-mass core.
{\it Icarus} {\bf 179}, 415-431.



\item Jacobson, R. A., Antreasian, P. G., Bordi, J. J., Criddle, K. E.,
Ionasescu, R., Jones, J. B., Mackenzie, R. A., Meek, M. C., Parcher, D.,
Pelletier, F. J., Owen, W. M., Roth, Jr., D. C., Roundhill, I. M., and J. R. 
Stauch 2006. The gravity field of the saturnian system from satellite
observations and spacecraft tracking data. {\it Astron. J.}, 
{\bf 132}, 2520-2526.

\item Jewitt, D., Chizmadia, L., Grimm, R., Prialnik, D., 2007.
Water in the Small Bodies of the Solar System.
In {\it Protostars and Planets V}. Eds., B. Reipurth, D. Jewitt, and 
K. Keil, University of Arizona Press, 863-878.

	
\item Johansen, A., Oishi, J. S., Low, M. M., Klahr, H., Henning, T., Youdin, A., 2007.
Rapid planetesimal formation in turbulent circumstellar disks	
{\it Nature} {\bf 448}, 1022-1025.

\item Johnson, T. V., and J. I. Lunine 2005. Density-derived
constraints on the origin of Saturn's moon Phoebe. {\it Nature},
{\bf 435}, 67-71.

\item Johnson T.V., Castillo-Rogez J. C., Matson D. L., Thomas P. C., 2009. 
Phoebe's shape: possible constraints on internal structure and origin. 
40th Lunar and Planetary Science Conference.







\item Kenyon, S. J., and Luu, J. X. 1999. Accretion in the early Kuiper Belt.
II. Fragmentation. {\it Astrophys. J.} {\bf 118}, 1101-1119.




\item Koponen, A., Kataja, M., Timonen, J., 1997.
Permeability and effective porosity of porous media.	
{\it Phys. Rev. E} {\bf 56}, 3319-3325.
	
\item Kretke, K. A., Lin, D. N. C., 2007.
Grain Retention and Formation of Planetesimals near the Snow Line 
in MRI-driven Turbulent Protoplanetary Disks. {\it Astrophys. J.}
{\bf 664}, L55-L58.








\item Lin, D. N. C., and J. Papaloizou 1993. On the tidal interaction between
protostellar disks and companions. In {\it Protostars and Planets III}
(E. H. Levy and J. I. Lunine, Eds.) pp. 749-836, Univ. of Arizona, Tucson.






\item Lunine. J. I.,
and Stevenson, D. J. 1982. Formation of the Galilean satellites
in a gaseous nebula. {\it Icarus} {\bf 52}, 14-39.
	
\item Lyra, W., Johansen, A., Zsom, A., Klahr, H., Piskunov, N., 2009.
Planet formation bursts at the borders of the dead zone in 2D numerical 
simulations of circumstellar disks. {\it Astron. \& Astrophys.} {\bf 497},
869-888.



\item McKinnon, W. B., Simonelli, D. P., and G. Schubert 1997. Composition,
internal structure, and thermal evolution of Pluto and Charon. In: Stern, A.,
Tholen, D. J. (Eds.), Pluto and Charon. Univ. of Arizona Press, Tucson, p.295.

\item Melosh, H. J., Nimmo, F., 2009.
An Intrusive Dike Origin for Iapetus' Enigmatic Ridge?	
40th Lunar and Planetary Science Conference.

\item Merk R., Prialnik D., 2006. Combined modeling of thermal evolution and 
accretion of trans-neptunian objects -- Occurrence of high temperatures and liquid water.
{\it Icarus} {\bf 183}, 283-295.



\item Morbidelli, A., Bottke, W., Nesvorny, D., Levison, H. F., 2009.
Asteroids Were Born Big. {\it Icarus}. In press.

\item Morbidelli, A., Levison, H. F., Gomes, R., 2008.
The Dynamical Structure of the Kuiper Belt and Its Primordial Origin.
In {\it The Solar System Beyond Neptune}. Eds., M. A. Barucci, H. Boehnhardt, 
D. P. Cruikshank, and A. Morbidelli, University of Arizona Press, 275-292.


\item Morfill G.E., Fechtig H., GrŸn E., and Goertz C.K., 1983. 
Some consequences of meteoroid impacts on 
SaturnÕs rings. {\it Icarus} {\bf 55}, 439-447.


\item Moses, J., 1992. Meteoroid ablation in Neptune's atmosphere.
{\it Icarus} {\bf 99}, 368-383.



\item Mosqueira, I., 2004. On Planet-Satellite Formation.
Presented at the KITP Program: Planet Formation: Terrestrial and Extra Solar, 
Kavli Institute for Theoretical Physics, University of California, Santa Barbara.

\item Mosqueira, I. and Estrada, P. R. 2003a. Formation of large regular
satellites of giant planets in an extended gaseous nebula. I. Subnebula
model and accretion of satellites. {\it Icarus} {\bf 163}, 198-231.

\item Mosqueira, I. and Estrada, P. R. 2003b. Formation of large regular
satellites of giant planets in an extended gaseous nebula. II. Satellite
migration and survival. {\it Icarus} {\bf 163}, 232-255.



\item Mosqueira, I., Estrada, P. R., 2005. On the origin of the Saturnian
satellite system: Did Iapetus form in-situ? 36th LPSC meeting, Houston, TX,
no. 1951.


\item Mosqueira, I., Estrada, P. R., and Turrini, D.,  2009.
Planetesimals and satellitesimals: the accretion of the regular satellites.
2nd ISSI-Europlanet Workshop Book. Submitted.

\item Multhaup, K., and T. Spohn 2007. Stagnant lid convection in the mid-sized icy
satellites of Saturn. {\it Icarus} {\bf 186}, 420-435.
	
\item Niemann, H. B., Atreya, S. K., Bauer, S. J., Carignan, G. R., Demick, J. E., Frost, R. L., 
Gautier, D., Haberman, J. A., Harpold, D. N., Hunten, D. M., Israel, G., Lunine, J. I., 
Kasprzak, W. T., Owen, T. C., Paulkovich, M., Raulin, F., Raaen, E., Way, S. H., 2005.
The abundances of constituents of Titan's atmosphere from the GCMS instrument on the 
Huygens probe. {\it Nature} {\bf 438}, 779-784.


\item Owen, T., Encrenaz, T., 2003.
Element Abundances and Isotope Ratios in the Giant Planets and Titan.
{\it Space Sci. Rev.} {\bf 106}, 121-138.

		
\item Person, M. J., Elliot, J. L., Gulbis, A. A. S., Pasachoff, J. M., Babcock, B. A., 
Souza, S. P., Gangestad, J., 2006. Charon's Radius and Density from the Combined 
Data Sets of the 2005 July 11 Occultation. {\it Astron. J.} {\bf 132}, 1575-1580.

\item Podolak, M., Pollack, J. B., Reynolds, R. T., 1988.
Interactions of planetesimals with protoplanetary atmospheres.
{\it Icarus} {\bf 73}, 163-179.

\item Pollack, J. B., A. S. Grossman, R. Moore, and H. C. Graboske, Jr. 1976.
The formation of Saturn's satellites and rings as influenced by Saturn's
contraction history. {\it Iacrus} {\bf 29}, 35-48. 


\item Pollack, J. B., O. Hubickyj, P. Bodenheimer, J. J. Lissauer,
M. Podolak, and Y. Greenzweig 1996. Formation of the giant planets
by concurrent accretion of solids and gas. {\it Icarus} {\bf 124}, 62-85.


\item Pollack, J. B., Podolak, M., Bodenheimer, P., and  B. Christofferson
1986. Planetesimal dissolution in the envelopes of the forming, giant planets.
{\it Icarus}, {\bf 67}, 409-443.

\item Pollack J. B., Reynolds R. T., 1974.
Implications of Jupiter's early contraction history for the composition 
of the Galilean satellites. {\it Icarus} {\bf 21}, 248-253.



\item Prialnik D., Podolak M., 1995. Radioactive heating of porous comet nuclei. 
{\it Icarus} {\bf 117}, 420-430.

\item Prialnik D., Sarid G., Rosenberg E. D., Merk R., 2008.
Thermal and chemical evolution of comet nuclei and Kuiper belt objects. 
{\it Space Sci. Rev.} {\bf 138}, 147-164.

\item Prinn R. G., Fegley B., 1981. Kinetic inhibition of $CO$ and $N_{2}$ reduction in 
circumplanetary nebulae: implications for satellite composition. {\it Astrophys. J.} 
{\bf 249}, 308-317.


\item Rafikov, R. R. 2002. Planet migration and gap formation by tidally
induced shocks. {\it Astrophys. J.} {\bf 572}, 566-579.





\item Ruskol, Y. L. 1975. Origin of the Moon. Origin of the moon NASA Transl. 
into english of the book ``Proiskhozhdeniye Luny'' Moscow, Nauka Press, 1975 
1-188.

\item Ryan, E. V., Davis, D. R., Giblin, I., 1999.
A Laboratory Impact Study of Simulated Edgeworth-Kuiper Belt Objects.
{\it Icarus} {\bf 142}, 56-62.



	
\item Scott, E. R. D., Krot, A. N., 2005.
Chondritic Meteorites and the High-Temperature Nebular Origins of Their Components.
In {\it Chondrites and the Protoplanetary Disk}., ASP Conference Series, {\bf 341}. 
Eds., A. N. Krot, E. R. D. Scott, and B. Reipurth. Astronomical Society of the Pacific, p.15.



\item Smoluchowski, R. 1981. Heat content and evolution of cometary nuclei. {\it Icarus}
{\bf 47}, 312-319.

\item Sohl, F., Spohn, T., Breuer, D., Nagel, K., 2002.
Implications from Galileo Observations on the Interior Structure and Chemistry of the Galilean Satellites.
{\it Icarus} {\bf 157}, 104-119.

\item Stern, S. A., and P. R. Weissman 2001.
Rapid collisional evolution of comets during the formation of the Oort cloud.
{\it Nature}, {\bf 409}, 589-591. 


\item Stevenson, D. J., A. W. Harris, and J. I. Lunine 1986. Origins of
satellites. In {\it Satellites} (J. A. Burns and M. S. Matthews, Eds.)
Univ. of Arizona Press, Tucson.



\item Svetsov, V. V., Nemtchinov, I. V., and Teterev, A. V., 1995. 
Disintegration of large meteoroids in Earth's atmosphere: Theoretical models. {\it Icarus}
{\bf 116}, 131-153.


\item Tamayo, D., Burns, J. A., Denk, T., 2009.
Dynamical Models for the Origin of Iapetus' Dark Material.
{\it BAAS} {\bf 41}.


\item Tanaka, H., T. Takeuchi, W. R. Ward 2002. Three-dimensional interaction
between a planet and an isothermal gaseous disk. I. Corotation and
Lindblad torques and planet migration. {\it Astropys. J.} {\bf 565}, 1257-1274.


	
\item Thomas, P. C., Veverka, J., Helfenstein, P., Porco, C., Burns, J., Denk, T.,
Turtle, E., Jacobson, R. A., 2006. Shapes of the Saturnian Icy Satellites
37th Annual Lunar and Planetary Science Conference.
	
\item Tremaine, S., Touma, J., Namouni, F., 2009.
Satellite Dynamics on the Laplace Surface. {\it Astron. J.} {\bf 137},
3706-3717.

\item Tsiganis K., Gomes R., Morbidelli A., Levison H. F., 2005. Origin of the orbital 
architecture of the giant planets of the Solar System. {\it Nature} {\bf 435}, 459-461.

\item Turner, N. J., Sano, T., Dziourkevitch, N., 2007. Turbulent 
Mixing and the Dead Zone in Protostellar Disks.
{\it Astrophys. J.} {\bf 659}, 729-737.

\item Ward, W. R., 1981.
Orbital inclination of Iapetus and the rotation of the Laplacian plane.
{\it Icarus} {\bf 46}, 97-107.


\item Weidenschilling, S. 1997. The origin of comets in the solar nebula: A
unified model. {\it Icarus}, {\bf 127}, 290-306.



\item Weidenschilling, S. J., 2000.
Formation of Planetesimals and Accretion of the Terrestrial Planets.
{\it Space Sci. Rev.} {\bf 92}, 295-310.

\item Weidenschilling, S. J., 2009.
How Big Were the First Planetesimals? Does Size Matter?	
40th Lunar and Planetary Science Conference.

\item Wetherill, G. W., and G. R. Stewart 1993. Formation of planetary 
embryos - Effects of fragmentation, low relative velocity, and independent 
variation of eccentricity and inclination. {\it Icarus}, {\bf 106}, 190. 

\item Youdin, A. N., and F. H. Shu 2002. Planetesimal formation by
gravitational instability. {\it Astrophys. J.} {\bf 580}, 494-505.

\item Young, E. D., 2001. The hydrology of carbonaceous chondrite parent
bodies and the evolution of planet progenitors. {\it Phil. Trans. R. Soc. Lond.} 
{\bf 359}, 2095-2110. 


\item Wong, M. H., Lunine, J. I., Atreya, S. K., Johnson, T., Mahaffy, P. R.,
Owen, T. C., Encrenaz, T., 2008. Oxygen and other volatiles in the giant
planets and their satellites. {\it Rev. Mineralogy and Geochem.} {\bf 68},
219-246.

\item Zahnle, K. J., 1992.
Airburst origin of dark shadows on Venus.
{\it J. Geophys. Res.} {\bf 97}, 243-255.
	
\item Zahnle, K., Mac Low, M., 1994.
The collision of Jupiter and Comet Shoemaker-Levy 9.
{\it Icarus} {\bf 108}, 1-17.
 

\end{list}

\end{document}